\documentclass[a4paper,12pt, authoryear]{elsarticle} 

\usepackage{setspace}

\usepackage{amsmath, amsthm, amssymb}

\usepackage{array}

\usepackage{natbib}

\usepackage{lscape}

\newcommand{\pd}[2]{\frac{\partial#1}{\partial#2}}
\newcommand{\brak}[1]{\left(#1\right)}
\newcommand{\tdt}[0]{_{t+\Delta t}}
\newcommand{\tmdt}[0]{_{t-\Delta t}}
\newcommand{\utf}[0]{\therefore}
\newcommand{\bmx}[1]{\begin{bmatrix}#1\end{bmatrix}}
\newcommand{\supscr}[1]{${^{\textrm{#1}}}$}

\begin{document}

\title{Which coordinate system for modelling path integration?}

\singlespacing

\author
{
    Robert J. Vickerstaff\supscr{a,b,*} and Allen Cheung\supscr{c} \\
    \tiny \supscr{*} Corresponding author. Email robert.vickerstaff@gmail.com\\
    \tiny \supscr{a}
    \tiny AgResearch Ltd, Lincoln Research Centre, Cnr Springs Road and Gerald Street,\\
    \tiny Private Bag 4749, Christchurch 8140, New Zealand. Tel +64 3 3218800\\
    \tiny \supscr{b}
    \tiny Department of Zoology, School of Biological Sciences, University of Canterbury,\\
    \tiny Private Bag 4800, Christchurch, New Zealand. Tel +64 3 3667001\\
    \tiny \supscr{c}
    \tiny The University of Queensland, Queensland Brain Institute \\
    \tiny / School of Information Technology and Electrical Engineering,\\
    \tiny Brisbane, Queensland 4072, Australia \\
}

\date{}

\maketitle


\section*{Abstract}
Path integration is a navigation strategy widely observed in nature where an animal maintains a running estimate, called the home vector, of its location during an excursion. Evidence suggests it is both ancient and ubiquitous in nature, and has been studied for over a century. In that time, canonical and neural network models have flourished, based on a wide range of assumptions, justifications and supporting data. Despite the importance of the phenomenon, consensus and unifying principles appear lacking. A fundamental issue is the neural representation of space needed for biological path integration. This paper presents a scheme to classify path integration systems on the basis of the way the home vector records and updates the spatial relationship between the animal and its home location. Four extended classes of coordinate systems are used to unify and review both canonical and neural network models of path integration, from the arthropod and mammalian literature. This scheme demonstrates analytical equivalence between models which may otherwise appear unrelated, and distinguishes between models which may superficially appear similar. A thorough analysis is carried out of the equational forms of important facets of path integration including updating, steering, searching and systematic errors, using each of the four coordinate systems. The type of available directional cue, namely allothetic or idiothetic, is also considered. It is shown that on balance, the class of home vectors which includes the geocentric Cartesian coordinate system, appears to be the most robust for biological systems. A key conclusion is that deducing computational structure from behavioural data alone will be difficult or impossible, at least in the absence of an analysis of random errors. Consequently it is likely that further theoretical insights into path integration will require an in-depth study of the effect of noise on the four classes of home vectors.

Keywords: navigation geocentric egocentric allothetic idiothetic

\section*{Introduction}
Path integration (PI) \citep{mittelstaedt80, mittelstaedt83} is a navigation strategy many animals are capable of using, including ants \citep{cheng09, muller88, ronacher08}, bees \citep{frisch67}, spiders \citep{moller94}, birds \citep{vonsaintpaul82}, crabs \citep{layne03, layne03a, zeil98}, rodents \citep{mittelstaedt80} and humans \citep{mittelstaedt91}, whereby the animal maintains an estimate of its location as it moves around, by integrating its velocity over time. The animal's estimate of its location is referred to as the home vector (HV), since it can be thought of as a vector connecting the animal's current location to the starting point of its journey, and allows it to return directly to the starting point. For reviews introducing this behaviour see \citet{collett00, gallistel90, redish99, wehner03}.

This paper undertakes a systematic comparison of several alternative ways of describing PI, based on the coordinate system in which the HV is expressed to define the spatial relationship between the animal and its home location. Previous mathematical models of PI have generally chosen a single coordinate system or frame of reference, often on the basis of incomplete or implicit assumptions. The only previous attempt to translate models between the alternative coordinate systems \citep{benhamou95} made a weakly justified assumption that PI should only be thought of using one specific reference frame. To complicate matters, most published neural network models of PI were not defined using the standard coordinate nomenclature. This has resulted in a wide range of seemingly inconsistent, even contradictory opinions concerning the necessary properties of a biological PI system. At various times, some version of each of the standard coordinate systems have been proposed to be the foundation of the correct model of PI in the arthropod literature.

In systematically comparing multiple coordinate systems, we explore and attempt to resolve several important questions. For example, the HV, being a vector, is suggestive of a polar representation. On the other hand, once near home, polar representations have significant drawbacks. An egocentric representation seems most intuitive for heading home. However, the egocentric position of home moves with every turn, even without displacement. Are these types of issues peculiar to specific coordinate systems? Do they persist following coordinate transforms? Can all major aspects of PI be represented equivalently in all coordinates systems? Are there theoretical clues as to which might be favoured biologically? It is important to recognise at the outset that there are intervening steps between the maintenance of a HV and its utilisation for finding home. For example, the HV could theoretically be represented in a geocentric framework (for definitions see below) but the motor commands needed to follow that HV may be more suitably transformed into an egocentric framework. This and related issues will be discussed later.

This paper covers the four standard coordinate systems within which PI has previously been modelled, and provides equations for translating HV values between them. It also introduces a scheme for classifying PI models into four extended families of coordinate systems for those models which do not easily fit into any of the four standard systems, allowing the classification of virtually any conceivable PI system. Based on the analytical results, a critique is given of the usage of particular coordinate systems in existing models. Equational models of biologically important aspects of PI are introduced in all four of the standard coordinate systems. These include HV updating, steering, searching, and systematic errors. This forms the basis for an objective analytical comparison of the properties of the coordinate systems. This paper considers PI in the absence of noise, and lays the foundation for later work which will incorporate the effects of random errors on PI to compare the noise tolerance of different classes of HV. Specifically, the definitions of the four extended families of coordinate systems were chosen to be suitable both for the present paper and for the study of noise tolerance.

The general conclusion reached is that, mathematically speaking, we confirm that all the coordinate systems can adequately and usefully \textit{describe} PI (in the sense of documenting or giving insights into navigation behaviour). However, geocentric Cartesian-like systems (see below for definition) appear the most robust solution for \textit{implementing} a full PI system (in the sense of modelling the way in which an animal's nervous system needs to process and update information), particularly when an allothetic compass is available.

\section*{Classification of Coordinate Systems for Path Integration}
Before considering which coordinate systems are most suitable for describing or implementing PI, it is essential to be able to classify the system which a given model uses. This section begins with four well known, standard coordinate schemes, and generalises them into four extended classes which can be used to classify virtually any conceivable model which can carry out accurate PI.

\begin{figure}
\centering
\includegraphics[width=10cm]{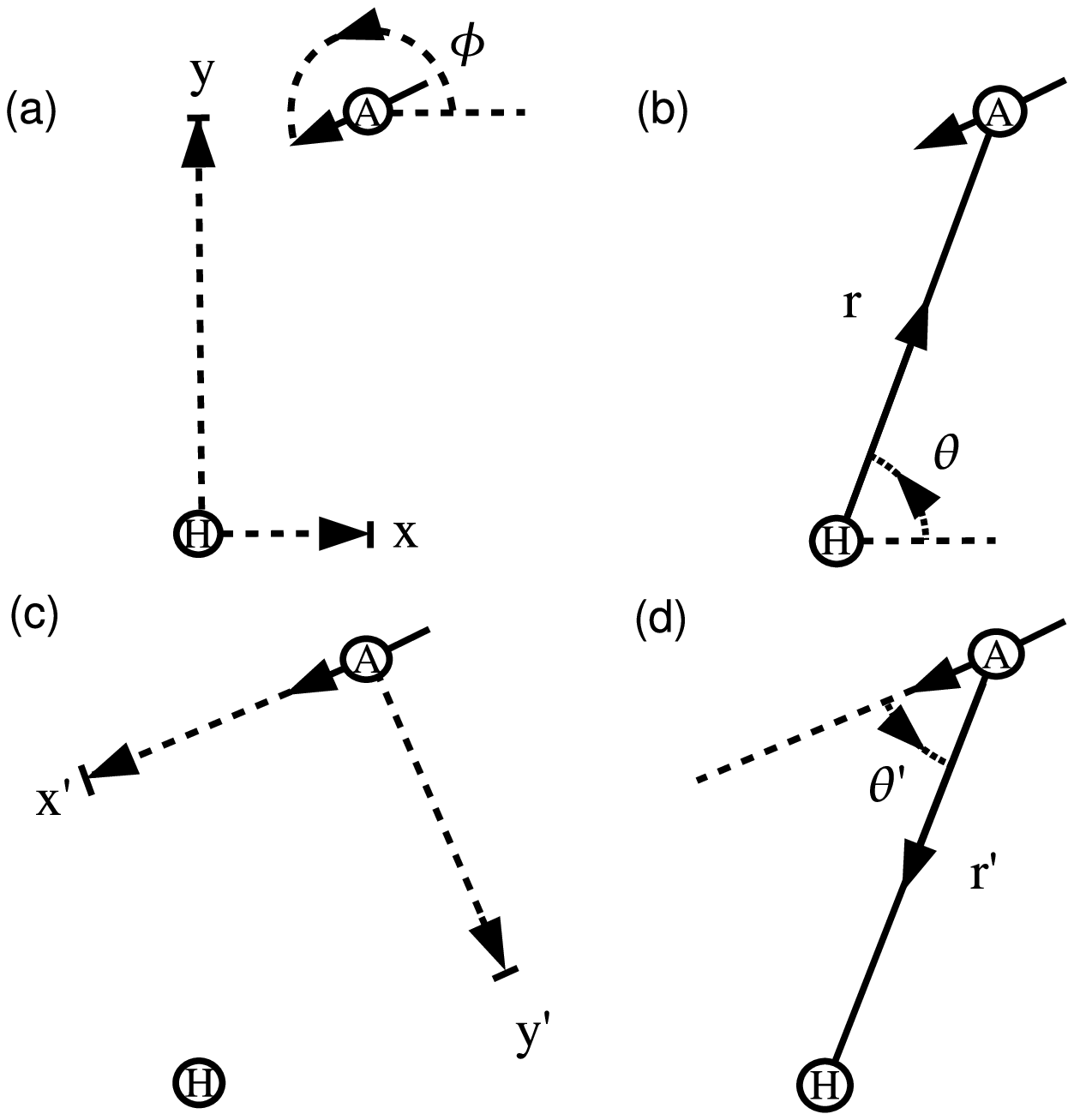}
\caption{Four ways to represent the same spatial relationship between animal and home. `A' is the animal's location, the associated arrow shows the orientation of its body axis. `H' shows the home location. Shown are HVs for each of the four `standard' coordinate systems considered in this paper. (a) geocentric Cartesian (GC), (b) geocentric polar (GP), (c) egocentric Cartesian (EC), (d) egocentric polar (EP). The angle $\phi$ (shown only for the GC case, but relevant for all cases) is the absolute compass heading, and is not part of the HV. The angles $\theta$, $\theta'$ and $\phi$ are measured positive in the anti-clockwise direction in radians.}
\label{fig:coord_sys}
\end{figure}

Most existing equational models of PI can be directly assigned to one of four `standard' coordinate systems according to the type of HV used. The class depends on whether the animal's position is given in Cartesian or polar form and whether the journey's starting point (the geocentric case) or the animal's body (the egocentric case) is used as the origin of the system. This leads to the four standard coordinate systems: geocentric Cartesian (GC), geocentric polar (GP), egocentric Cartesian (EC) and egocentric polar (EP). This paper considers only PI on a flat two dimensional plane, hence the simplest complete HV contains two values. Such a system can be extended to include PI on a non-flat surface by taking account of the local gradient, without the need for a full three dimensional HV. Desert ants appear to use such an approach, where an essentially two dimensional PI system is made to cope with uneven ground \citep{grah05, grah07}.  This paper will use the following symbols to express the four standard types of HV: GC as $(x,y)$, GP as $(r,\theta)$, EC as $(x',y')$ and EP as $(r',\theta')$. The symbol $\phi$ will be used to indicate the animal's compass heading measured anti-clockwise from the direction of the x axis. Fig.\ \ref{fig:coord_sys} shows the meaning of the four types of HV diagrammatically. Table 1 gives the equations needed for converting any standard HV into any other. Table 2 summarises all the abbreviations used in this paper.

Geocentric coordinates express the animal's position relative to the ground, with the origin corresponding to the starting point of the outbound journey and the direction of the axes corresponding to fixed directions with respect to the ground. The geocentric frame of reference is a special case of an `allocentric' or `exocentric' one: allo- or exocentric coordinates are those defined relative to something external to the animal's body, but in this paper the external reference used is always assumed to be the ground (for example the model animal is never on a table-top which can be rotated relative to the ground), making geo- synonymous with allo- and exocentric. Egocentric coordinates express the home position relative to the animal's current position and orientation. The origin in this case is the centre of the animal's body, the $x'$ axis corresponds to the forward direction along its body axis and the $y'$ axis to a perpendicular axis pointing to the left from the animal's point of view.

This paper presents equational PI models from the four standard classes of coordinate systems, because they encapsulate the essence of the majority of existing models, and they are mathematically convenient for analysis. However, to properly classify all the neural network models reviewed below (except \citet{jander57}), it is necessary to broaden each of the standard coordinate systems into an `extended' class of coordinate systems based on the vectorial scheme employed by \citet{cheung07a}. The four extended classes are: geocentric static-vectorial (GS, the class containing GC); geocentric dynamic-vectorial (GD, containing GP); egocentric static-vectorial (ES, containing EC); and egocentric dynamic-vectorial (ED, containing EP) representations (abbreviations summarised in Table 2). The distinction between geo- and egocentric remains the same for the extended classes but we introduce the concept of static and dynamic vectors. Static vectors are a superset of Cartesian vectors, where the HV representation now contains two or more values which are all distances measured in strictly fixed (static) directions within a geocentric or egocentric space. For example instead of measuring distances along $x'$ and $y'$ axes defined relative to the animal's body and separated by 90 degrees (a standard EC vector), we can measure three distances along axes defined relative to the animal's body and separated by 60 degrees. This is then an egocentric static vectorial representation and a model using such a HV would be classed as ES. If instead one or more of the distances are measured along an axis whose direction (within the defined ego- or geocentric space) is dynamically updated with changes in heading, then the vector is dynamic. This is clearly the case for all polar coordinates, since the distance defined by the scalar coordinate, such as $r$, is in a direction defined by the angular coordinate, such as $\theta$. It is clear that some key properties of the standard coordinate systems must also hold for their respective extended classes. A comprehensive demonstration of this is, however, beyond the scope of the present paper (but see Appendices C and D for a brief examination of two general classes of GS HV update equations).

One further definition will now be introduced, relating to the nature of the directional cues available to the PI system. While not part of the classification of the HV, the nature of these cues has important consequences for the information processing required by the PI system. In the present account directional cues will be classified as allothetic or idiothetic following the scheme presented in \citet[][Table 1]{cheung08}, depending on how the animal maintains an estimate of absolute heading. If the animal is required to perform neural integration to obtain the estimate of heading, then the directional cue is idiothetic. If no integration is required and the cue indicates absolute heading directly then the cue is allothetic. For example, rotational optic flow is generated from `allothetic' sensory data (the movement of the eye relative to objects in the external environment), but provides only a measure of rotation rate, which must be integrated internally to estimate absolute heading, and is therefore classified here as an idiothetic directional cue. This definition differs from that used by \citet{mittelstaedt00}, who defined idiothetic as ``spatial information that can \textit{only} be gained by means of the agent's active or passive movement'', thereby excluding optic flow, which could arise from the flow of objects in air or water past a stationary animal. However, since rotational optic flow accumulates errors in an analogous way to (idiothetic) vestibular or proprioceptive cues, we adopt the definition of \citet{cheung08}, since it is a more convenient basis for the analysis of errors in PI, and still agrees with Mittelstaedt's definition in spirit.

The geometrically correct methods for updating a HV in all four of the standard coordinate systems are given in Table 3 (HV Updating column), presented as continuous-time differential equations. These equations show how to update geocentric HVs using allothetic directional cues and egocentric HVs using idiothetic directional cues, since these are the most concise forms. For comparison and completeness, Tables 4 and 5 show the continuous-time equations required to update HVs using the alternative kind of compass input, and also present geometrically correct equations in discrete-time for all cases.

\section*{Coordinate Systems in Existing Models}
The choice of coordinate system must be of fundamental importance for understanding animal PI, since it determines the neural architecture, the nature of the information processing, and perhaps the type of sensory input required to perform PI-related behaviours. However, it is rare to find a complete discussion of the subject in published PI models. Typically, a published model adopts one particular coordinate system without reviewing the alternatives. Therefore, one might consider many of the existing models to be proofs of concept or existence proofs, where such a review of alternatives could be considered as beyond the scope of the study. One obvious difficulty in the arthropod literature is the lack of neurophysiological data, particularly in behaving animals. The paucity of data does allow room for speculative modelling. However, an existence proof cannot tell us whether behavioural consequences are specific to the assumed PI coordinate system. In contrast, the simple underlying geometrical structure of PI (limited admittedly by its noisy and approximate biological implementation) leads to simple methods for translating from one coordinate system to another (see Comparison of Coordinate Systems). Hence we should expect to be able to think about and describe a PI system from different reference frames, even if its actual neural implementation is in one specific system. Furthermore, a consideration of PI in alternative reference frames may lead to insights not obvious in the original coordinate system.

\citet[][in German, summarised in English in \citet{maurer95}]{jander57} introduced the first published mathematical model of PI. The model assumed that the animal moved at a constant speed, making it simpler than most subsequent models, which generally provided some means to accommodate variable speeds. It also contained no true HV. The animal instead maintained a time-weighted average of its heading relative to an external light source, giving a single scalar in place of the usual two valued HV. The scalar was effectively the angular coordinate of a GP HV. Such a choice was natural since, during any initial straight part of a foraging trip where the home location remains directly behind the animal, the angle to the light is equal to the angular part of a GP HV, but is unrelated to the angular part of an EP HV which is always $\pi$ during this period. Unfortunately, although presented as being correct, Jander's model is geometrically erroneous \citep{maurer95} and, although clearly related to GP PI systems, is not strictly a member of this class.

The Mittelstaedts' PI model, the bicomponent model, \citep{mittelstaedt73, mittelstaedt83} used a GC HV. This choice seemed to stem from earlier work related to orientation, including the work of von Holst on orientation in fish \citep{mittelstaedt83a}. Fish motor responses to deviations from an external orientation cue were found to be approximately sinusoidal, thus providing feedback capable of stabilising the animal. A problem arose when considering how an animal may orient at some arbitrary angle to an external cue. Mittelstaedt proposed his reciprocal modulation of bicomponents theory \citep{mittelstaedt62} as the solution to this problem. The present and desired angular deviations from the external cue were each broken down into a sine and cosine component (hence `bicomponent model'). The two pairs were `reciprocally modulated' (the sine of one pair being multiplied with the cosine of the other pair and vice versa), then the values were summed, and used to control the rotation of the animal. For PI, the sine and cosine values from an allothetic compass cue (or sometimes an idiothetic compass generated from integrating rotation rate) were used as inputs. A GC HV was the natural choice, obtained by multiplying the compass inputs by the forward speed and integrating over time (see Table 3, GC HV Updating). Homing was performed using reciprocal modulation, the system deriving its desired compass heading from the current HV (see Table 3, GC Homing). GC coordinates therefore stemmed indirectly from the observation of sinusoidal motor responses in animals, generalised using the reciprocal modulation hypothesis. Neurons showing actual sine and cosine response functions to the orientation of external stimuli have since been found in nature: in response to touch in leeches \citep{lewis98} and air flow direction in crickets \citep{miller91}. The latter could in principal constitute an allothetic compass cue given a sufficiently stable wind direction. Skylight cues are generally considered to be the most important allothetic compass cue for insects, and here no simple sine wave responses are known. Instead, polarisation-sensitive (POL) neurons in the early stages of the polarisation vision pathway of crickets give a sinusoidal response with a period of 180$^\circ$ \citep{henze07} rather than the 360$^\circ$ predicted by the bicomponent model. There is evidence that information from polarisation and spectral cues as well as the solar elevation are combined to produce a reliable, time compensated compass signal, but as yet no explicit sine-wave signal has been detected \citep{pfeiffer07}. A positive feature of the bicomponent model was its explicit inclusion, often lacking in later models, of all the information processing required to generate a HV from clearly defined compass inputs and to generate motor control signals from the HV capable of taking the animal back home.

\citet{muller88} used a PI model based on a GP HV, to explain PI in desert ants. This choice probably stemmed from the model's starting point (drawing inspiration from Jander's model) as a calculation of the arithmetic mean of the animal's compass heading weighted by the distance walked in each direction. The use of Jander's model as the starting point, rather than the Mittelstaedts' model, was motivated by the idea that the ant was performing a computation that is mathematically simpler than the geometrically correct one \citep{maurer95}. In spite of its starting point, the final form of the model is simple to derive from the discrete-time approximation (see Table 4, GP Discrete) of the geometrically correct continuous-time GP equations (see Table 3, GP HV Updating) by using a piece-wise linear approximation of the cosine function, a polynomial approximation of the sine function, and an additional scaling factor. These complications weakened the apparent link to Jander's GP-like model, and made the choice of coordinate system seem more arbitrary. Observed desert ant PI behaviour suggests a potentially disastrous drawback of using polar coordinates. The ant's HV is assumed to be set to zero magnitude ($r=0$) when the animal initially leaves home to begin foraging. During homing the animal is assumed to be aiming to reach the goal location indicated by the HV (the location where its HV will have reached zero magnitude again), and during search behaviour it repeatedly passes over this location \citep{wehner81}. The polar equations produce an infinite rate of change in the angular HV component at this location, due to the division by $r$ (or $r'$ for the egocentric equations, see Table 3, GP and EP HV Updating) and more generally, the closer to home the animal gets the larger the changes in the angular component must become. The Hartmann-Wehner model \citep{hartmann95} was a neural network implementation based on polar coordinates (see below). One important advantage of the M\"uller-Wehner model was that its parameters could be adjusted to reproduce the observed systematic deviations from direct homing when desert ants were forced to walk along L-shaped channels \citep{muller88} (see below for a more detailed discussion of this phenomenon). However, it remains uncertain how generally this behaviour relates to natural navigation. Furthermore, as alluded to previously, it is not clear whether this ability of the model supports the argument for a GP HV for biological PI since it was not shown whether a translation of the model into other coordinate systems would alter its ability to reproduce the errors. Finally, this model is not the only one able to account for the errors \citep{merkle06, vickerstaff05}.

\citet{gallistel90} presented two PI models in the form of equations and flowchart diagrams. As a prelude, Gallistel discussed the need to choose between polar and Cartesian coordinates for modelling PI, both apparently of the geocentric kind (``Does the animal represent its position relative to the nest on a rectangular grid or as a certain distance and angle away from the nest?'' p.71). Gallistel considered polar coordinates to be appealing because they allow the HV to be stored in a form closer to that required for homing, but that they were less computationally attractive due to the presence of positive feedback loops in the updating system (although no quantitative treatment of this claimed weakness was given). The discussion lacked any explicit reference to the choice between geocentric and egocentric coordinates. The first of the two PI models presented used GC coordinates and was similar to the Mittelstaedt model, except that it used a more complicated homing system. The assumption appeared to be that the PI system must output the distance and turning angle required to reach home from the current location. To find these the system used a square root function and additional angular calculations. The Mittelstaedt model had already clearly shown how to accomplish homing without these extra complications, using the reciprocal modulation method (see Table 3, GC Homing). The second of the models presented was a polar PI system. From the definition of the symbols used, the HV was intended to be of the EP kind, but the update equations given were not correct for this kind of HV \citep[as][have observed]{benhamou95}. To be correct GP update equations, both right-hand sides would need to be multiplied by negative one.

\citet{benhamou90} discussed the effects of noise on PI, which was referred to as `the egocentric coding process'. The authors concluded that PI systems are much more sensitive to noise from idiothetic directional cues than either noise from allothetic compass cues or noise from estimating forward speed. The introduction set out a dichotomy between exocentric and egocentric coding processes. An exocentric coding process meant landmark-based navigation, where goal locations were memorised relative to landmarks. An egocentric coding process consisted of processing `route-based' (velocity cue) information, and was stated to be largely synonymous with the terms dead-reckoning, inertial navigation, path integration and route-based navigation as used by other authors. The authors were therefore distinguishing locations memorised in geocentric and egocentric terms. PI was assumed to function using an EP representation of one or more goals (hence `egocentric  coding process'), justified by the supposed lack of dependence between individual stored goal coordinates in this system. The idea was to avoid positing a map-like representation of the animal's entire home range. However, if two or more EP goal coordinates were updated using the same velocity cues, they must lie within the same coordinate system, and therefore be dependent in this sense. Yet, given that egocentric coordinates must be constantly updated as an animal moves, the relative positions of two landmarks may diverge through cumulative errors thereby reducing their spatial relationship in representational space. This is certainly not a positive feature of EP coordinates for PI. Conversely, geocentric (exocentric) coordinates do not automatically lead to a cognitive map. \citet{collett04} argue that if an animal can associate PI coordinates to an array of visually defined places, then by this fact it possesses a cognitive map, which Benhamou et al.\ wish to avoid. Evidence from desert ants suggests they cannot perform this kind of association \citep{collett03}, but this clearly doesn't invalidate the many geocentric PI models for this animal, which do not rely on the association of HVs with landmarks. In conclusion, the justification given by Benhamou et al.\ for using egocentric coordinates was weak, whereas the choice of polar coordinates was made without any clear justification.

\citet{wittmann95} presented a neural network model of PI where the HV was stored on an array of neurons called a sinusoidal array. This design originated from hypotheses about how rats might encode the distance and direction to landmarks, but had the distinct advantage of making vectorial addition rather trivial, and it was also noise tolerant. The activity pattern along the array described a sine wave whose wavelength spanned exactly 360 degrees of angular space, and whose amplitude and phase encoded the distance and angle of a geocentric HV respectively. At first glance, this model resembles a polar representation. However, for present purposes the model must be classified as GS (i.e.\ a superset of GC). To understand this, we need to consider the HV updating process. Each step was represented as a sinusoidal activity pattern, and the net HV was generated from the cumulative sum of the activity patterns of all steps taken. Each neuron in the array represents a fixed geocentric compass direction. Any two neurons separated by one quarter of the length of the sinusoidal array could be interpreted as GC coordinates on a certain set of fixed geocentric axes. The activity level and location of the most excited neurons in the array indicated the linear and angular component of the HV in GP form, without having to perform anything resembling a square root or arctangent function, \citep[cf.][]{gallistel90}. Hence the information in the HV was essentially available simultaneously in Cartesian and polar forms. The use of a geocentric rather than an egocentric HV arose naturally from the use of an array of input neurons representing an allothetic compass cue. The neurons of the sinusoidal array HV then essentially accumulated input from the corresponding geocentric compass input directions. The apparent contradiction of a GP HV readout from a GS representation stems from using redundant properties within the one PI system. In a GS representation, accurate PI can occur in at least two different ways. For a small number of static vectors, accurate PI requires a particular transformation of the heading direction prior to updating the HV (see Appendix C). However, for a large number of static vectors, the transformation is irrelevant (see Appendix D). \citet{wittmann95} used a large array of neurons and also assumed a sinusoidal activity pattern simultaneously, thereby carrying redundant information about the HV.

\citet{hartmann95} presented a neural network model of PI which set out to reproduce the systematic homing errors seen in desert ants in specific experimental conditions, taking the M\"uller-Wehner model \citep{muller88} as a starting point. However, instead of deriving the model from a distance-weighted mean-heading model as \citet{muller88} did, it was introduced as a modification of geometrically correct GP PI \citep[see][Eqns 4 and 5]{hartmann95}. Although this foundation contained minor inaccuracies (first, the paper dealt exclusively with angular units of degrees but Eqn.\ 5 was correct only for radians; second, the equations given were in fact only approximations to the geometrically correct discrete-time PI equations, see Table 4), these would probably not have prevented the model from performing PI quite accurately under most conditions. The neural implementation of these equations was then made deliberately imprecise, in a manner which mostly mimicked the mathematical approximations employed by the M\"uller-Wehner model, in order to reproduce the systematic homing errors reported by \citet{muller88}. In the neural network of \citet{hartmann95}, the HV was explicitly separated into magnitude and angular components in the manner of polar coordinates. The distance magnitude was in the direction indicated by the variable angular component, hence the model must be classified as GD, a clear consequence of being based on the GP M\"uller-Wehner equational model. The HV magnitude was represented by how far activation reached along a linear chain of neurons, and the HV direction was represented by the position of a patch of activity on a circular chain of neurons. Unlike the Wittmann-Schwegler model, information related to the Cartesian translation of the HV was not simultaneously available within the network using this representation. This direct implementation of a polar-like HV appears to inherit the divide-by-zero feature of the M\"uller-Wehner model, requiring an infinitely fast rate of change of the angular HV component at the nest. The authors assumed that this value would change only very slowly ``as long as the ant does not run around its nest'' \citep[][p.488]{hartmann95}, but in fact the problem would be if the ant ran \textit{over} its estimate of the nest location, a behaviour which is observed during ant search behaviour \citep{wehner81}. The activity patch representing the HV angular component would be required to jump discontinuously by a half turn, which seems impossible given the shifter circuitry employed by the model. \citet{chapman98} reported problems getting the model to function correctly during tests on a mobile robot, particularly during homing, but did not state that this problem was limited to the vicinity of the nest, suggesting perhaps a further but as yet poorly defined problem of this model. One possibility may be the fact that approximate update equations can lead to negative values of magnitude which cannot be sustained by the aforementioned neural chain architecture.

\citet{benhamou95} studied five PI models with the aim of translating them into discrete time recurrent EP formulas. It is clear that the authors considered the EP reference frame to be the most appropriate for models of PI, but this choice was not adequately justified.
The rationale given was mostly tautological: ``From the animal's point of view, memorising the home location... is likely to be an egocentric coding process'' (p.\ 463). Two references were given to papers which suggest that the caudate nucleus may be involved in processing egocentric positional information. This is weak justification since this brain area may not be involved in PI (or at least may not be the core of PI), and in any case three of the five models considered were originally formulated for arthropods which lack this brain area all together. Despite the weak basis for choosing the EP coordinate system, the authors felt justified in dismissing the bicomponent model \citep{mittelstaedt73, mittelstaedt83} on the grounds that it cannot be easily converted into EP according to their methodology.

\citet{benhamou97a} considered systems of reference for navigation in general, including PI. He noted that geo- and egocentric formulations of PI are in some sense mathematically equivalent, but claimed that egocentric formulations do not require an external reference direction whereas geocentric ones do. This is incorrect. It is possible to store the HV in geocentric terms but to employ idiothetic directional cues. For example the Hartmann-Wehner model \citep{hartmann95} used this approach, as did a version of the bicomponent model \citep{mittelstaedt85}. An egocentric HV can also use information from an allothetic compass. Tables 4 and 5 give details of how either kind of directional cue can be used with either kind of PI system. Even if the claim of \citet{benhamou97a} was true, it could not provide much justification for using an egocentric reference frame since \citet{cheung07, cheung08} have shown clearly that an external compass leads to more accurate navigation than using purely idiothetic cues. Benhamou stated that when considering the neurobiological basis of PI ``it seems better to adopt the animal's point of view, and hence to consider that path integration involves an egocentric localisation system'' \citep[][p.151]{benhamou97a}. \citet{wehner96} is cited in support of this view, despite the fact that the latter refers to the model suggested by \citet{muller88}, who presented a geocentric PI model.

\citet{maurer98} presented a neural network PI model employing an EP-like HV. The magnitude and directional components of the HV were stored in separate neural structures, hence the HV can be classified as ED (since it contained more than two values representing the polar HV, it was not strictly a simple EP HV). The magnitude was indicated by the position of an activity packet on a linear array of neurons and the direction by the position of an activity packet on a circular array of neurons. The network was trained using back propagation and had no homing mechanism - the output of the model was simply the HV itself. Unlike the model of \citet{hartmann95}, the angular component of the HV was not moved by dedicated shifter circuitry, and may have been able to meet the need for this part of the HV to jump instantaneously by a half turn were the animal to pass over the estimated home location (although this ability was not tested). This network model was compared with one which (rather like Jander's model) did not contain a representation of the HV magnitude, consisting of only the angular component of the first model.

\citet{kim00} outlined a neural-network based PI system. The HV was stored on a circular array of neurons, each corresponding to a fixed compass heading (or rather to a small contiguous range of compass headings), hence the model was GS. When the animal was travelling in a direction corresponding to a given array neuron, it was assumed to accumulate activity in proportion to the distance the animal travelled in that direction. The mechanism mapping directional and speed inputs into this integrative behaviour of the array neurons was not specified. Unlike the Wittmann-Schwegler model a neuron's activation level did not decrease when the animal was heading opposite to its preferred direction, meaning that the state of the HV neurons actually retained some information about the path taken to reach a given location, in addition to the location itself. Also, since the activity pattern was not sinusoidal, there was no redundant GP information in the HV representation (see overview of Wittmann-Schwegler model above). Making the same apparently unnecessary assumption (see Table 3 GC Homing) as \citet{gallistel90} the authors indicated that they assume the HV must have a read-out mechanism providing the distance and direction to home in polar form. In line with this a second circular array of neurons was proposed which processed the HV stored in the first array to yield a single active neuron pointing in the direction of home, with the magnitude of activation indicating the distance. The way in which this signal was to control motor output was specified only using a fragment of pseudocode, rather than neural mechanisms.

\citet{biegler00} classified PI systems with respect to their level of capability, such as whether they could store more than one goal location, and how landmark-based positional fixes could be used to increase the accuracy of navigation. Coordinate systems were covered briefly, including a classification of HVs into polar, Cartesian and map-like, and as egocentric or exocentric. Biegler's classification of map-like HVs as a class by itself contrasts with the approach taken here, where they will be classified as GS. This choice is motivated by the fact that fixed directions within the abstract map-like sheet of cells in map-based PI models correspond to fixed directions in the geocentric reference frame. In general agreement with the present account, Biegler considered all types of HV representation as strictly mathematically equivalent.

\citet{vickerstaff05} produced a neural network model of PI using a genetic algorithm (a system for producing artificial evolution \citep{goldberg89}). The best evolved network obtained was found to be closely related to the Mittelstaedt PI model when analysed, and used a GC HV. This was the only type of HV encountered in the various runs of the genetic algorithm. The failure of evolution to utilise other forms of HV probably stemmed from the choice of the compass inputs made available to the networks, which were either of the exact kind or similar to those required for the GC HV updating equations of the Mittelstaedts' model in most of the experiments. Unfortunately, there were no experiments which used rotation rate sensors in place of compass sensors. The equational form of the egocentric HV update process appear simplest when rotation rate $\dot\phi$ is the directional cue, while geocentric update equations appear simplest when an absolute compass heading $\phi$ is available (see Table 3, HV Updating). Hence one may not have expected an egocentric HV to be easily evolved under the evolutionary circumstances modelled. In detail, there may also be some doubt concerning the biological plausibility of the evolved neural network due to the use of a non-standard neuron feature which allowed the network to perform multiplications more easily. However, for comparison, both \citet{hartmann95} and \citet{wittmann95} included an \textit{ad hoc} multiplication-like mechanism in their network models to accommodate the effects of variations in the animal's speed into the HV updating process. The neuron-model \citep{beer92} used in the neural network model of \citet{vickerstaff05} was based on a leaky integrator equation, so that the HV representation decayed continuously. Due to the artificially imposed selection of efficient homing, the network evolved to compensate for the decay of the HV so that no systematic navigation errors occurred. When this compensation process was disabled manually, the resulting systematic navigation errors were found to closely resemble those reported in desert ants under the specific experimental conditions tested and reported by \citet{muller88}. The network also evolved to perform a kind of searching behaviour when the simulated agent had returned to the vicinity of, but had not yet found, the nest. The features of a HV stored on leaky integrators, and of the searching behaviour displayed by the model will be considered in detail in the next section.

\citet{merkle06} presented an equational PI model using an EC HV. The authors first reviewed existing models with respect to their choice of coordinate system, and noted that EC had previously been neglected. The simple form of the EC HV updating and homing equations (Table 3) were seen as an advantage, as was the use of egocentric coordinates, because the animal ``perceives all sensory inputs relative to its own position and orientation'' (p.\ 388, \textit{ibid}). While this is certainly true, it does not necessarily follow that egocentric coordinates are the natural choice for PI. According to the geometrically derived PI equations considered in the present paper (Table 3, HV Updating) an allothetic compass cue (denoted $\phi$ in the equations) giving the animal its absolute heading, is more directly relevant for updating a geocentric than an egocentric HV. The rate of change of the absolute heading would have to be calculated in order to update an egocentric HV (see Table 5), implying a more complex computation. The equations show that the perception of its own forward speed ($s$ in the equations) is equally relevant to ego- and geocentric PI. The perception of rotation rate ($\dot \phi$) appears more relevant to ego- than geocentric PI, since the rate must be integrated to estimate absolute heading before being used to update a geocentric HV (see Table 4), but has been shown to be a much less reliable cue than an allothetic compass \citep{benhamou90, cheung07, cheung08}. A further potential disadvantage of using EC coordinates is that the HV must be updated in response to rotation of the animal as well as translation. The further the animal is from the home location the greater the magnitude of the changes caused by rotation: a full turn of the animal causes the HV value to perform a complete circle in representational space. If rapid changes to the HV have an energetic cost or are associated with an increase in errors then an egocentric HV would be disadvantaged for this reason alone. This line of argument predicts that an animal using an egocentric HV might find it advantageous to limit it rotations, especially when far from the nest, but that one using a geocentric HV would not have any such incentive on this account. We save a more detailed treatment of this line of argument for future work, but note that desert ants certainly follow tortuous paths during foraging, including when they are far from the nest \citep{wehner03a}. \citet{wehner81} also found a high maximum rotation rate (during nest search) in \textit{Cataglyphis albicans} ants of $4000^\circ$ per second. Merkle et al.\ undertook an investigation of a number of possible explanations of systematic homing errors in desert ants, and concluded that (amongst other candidate models) the use of leaky integration with their EC HV could explain the data well. As the present paper will show, the use of leaky integration in any of the four standard coordinate systems produces exactly the same error patterns. Therefore this fact does not lend much support to the hypothesis that EC coordinates are being used by these species. Other plausible explanations of PI systematic errors may yet arise and it remains to be seen whether they may be equivalently represented in all four coordinate systems.

\citet{haferlach07} presented a genetic-algorithm evolved neural network model of PI, which used a GS HV. They provided the network with the same type of compass sensors as the Vickerstaff-DiPaolo model. Unlike \citet{vickerstaff05}, the network topology was constrained during evolution to force each compass sensor to have its own directly attached integrative neuron. Both of these features probably helped predispose the system to find a GS solution - the genetic algorithm found the simplest solution, namely each integrative neuron integrated the output of its compass sensor. Since their simulated agent travelled at a fixed speed, this is sufficient to provide a GS HV (i.e.\ no input from a speed sensor is required). As the authors note the network did not store magnitude and angular HV information separately, and is only superficially similar to the Hartmann-Wehner model in this sense. The main differences from the Vickerstaff-DiPaolo model with respect to HV updating are that the network was given three compass inputs instead of two, leading to a three part GS HV rather than a two part strictly GC HV, and like Jander's model, the simulated animal is assumed to move at a fixed speed. The homing mechanism also differed: whereas the Vickerstaff-DiPaolo model exploited the `reciprocal modulation' solution proposed by Mittelstaedt \citep{mittelstaedt62}, the Haferlach et al.\ model used a method similar to that in the Wittmann-Schwegler and Hartmann-Wehner models, and seems to have avoided the need for an explicit multiplication operation in the homing mechanism. This suggests an hypothesis about the problem Mittelstaedt was solving with reciprocal modulation: if the HV is expanded into a redundant encoding containing more than two values (GS rather than the subset GC), a neural network can produce the correct homing direction using a simple repulsion-like mechanism (where the animal's heading is repelled from the direction of maximum HV magnitude) without needing to implement reciprocal modulation (i.e.\ multiplication). Vickerstaff-DiPaolo used a two value HV and appeared to require reciprocal modulation to generate direct homing paths from all positions. In their first model, \citet{haferlach07} used a three-valued HV and appeared to generate direct homing trajectories without multiplication (direct homing was explicitly selected for during evolution through the genetic algorithm's fitness function).

\citet{bernardet08} presented a neural network model of PI using a GS HV where a population of neurons was divided into 36 groups each representing a fixed compass direction. The number of active neurons in a group recorded how far the animal had travelled in this direction. Like the Kim-Hallam model, the activity of HV neurons could only increase during motion i.e., motion in the opposite direction did not decrease activation levels. The model appears to be unique in that it included a mechanism for updating the HV for variable speeds of travel which did not require neurons to perform multiplication. Rather than having compass neurons with graded firing rates, compass neurons were either fully active or silent, so direction was represented by the location of a single active neuron in a circular chain. The speed was encoded in the firing rate of a `velocity' (\textit{sic}) neuron. A set of gater neurons passed the speed signal on to the HV memory neurons only for the memory group corresponding to the direction of the active compass neuron. No true homing mechanism was presented. Instead the authors, like \citet{gallistel90} and \citet{kim00}, assumed that the HV memory neurons must provide a `read out' mechanism that outputs the direction and magnitude of the HV.

For completeness, we briefly mention mammalian PI models. The hippocampal place cells \citep{okeefe71}, medial entorhinal grid cells \citep{hafting05} and the more diffuse head direction cell network \citep{taube90a, taube90b} studied in rodents have been hypothesized to form the basis of a PI system \citep{burak09, burgess07a, mcnaughton96, mcnaughton06, samsonovich97}. Both the toroidal neural network model, reviewed recently by \citet{mcnaughton06}, and the algorithmic oscillatory interference model, recently reviewed by \citet{burgess08} use map-like PI schemes in a geocentric reference frame. There is rapidly growing consensus that rodent PI HV updating occurs in the grid cell networks, which typically display geocentric firing patterns as rats moved about experimental environments. The rodent PI models differ from the arthropod PI models presented above in using an abstract two dimensional space, or chart, within which neural activity patterns formed one or more activity packets, the location and movements of which fulfilled the role of the HV. This obvious difference does not however put them beyond the scope of our classification scheme: using the extended coordinate system classification scheme, neural units which fire in response to being at specific positions in space are representing GS vectors. Map-based PI models are therefore a special case of GS PI where the static vectors have fixed length as well as geocentric direction, hence limiting their spatial receptive field to a point (or small region). A 2-D array of such neural units could theoretically cover the entire PI range with or without tessellation. Although not so far proposed in the literature, we note here that an ES HV could also be implemented using a map-like scheme, where the activity packet position represented the nest's location relative to the current position and orientation of the animal. Thus map-based PI systems fit easily within the classification scheme introduced in the present account.

\section*{Comparison of Coordinate-Systems}
This section considers several biologically important aspects of PI (see Table 3): HV updating, systematic PI-navigation errors \citep{muller88, sommer04}, PI-mediated homing and PI-mediated systematic searching \citep{muller94, wehner81}. These aspects of PI are used as the basis for comparing the different coordinate systems. Systematic errors and systematic searching are considered in some detail. Differential equations defining a continuous-time model of each of these aspects of PI (Table 3) are used as the main basis for discussion, for reference discrete-time versions of HV updating, homing and searching are also given (Tables 4, 5, 6). Each of the models is presented in all four of the standard coordinate systems, and in each case the four models were obtained through exact transformation of the GC version of the model into the other three reference frames, ensuring that the four are mathematically strictly equivalent. The four sets of equations for each coordinate system were also validated using numerical integration \citep[Runge-Kutta fourth order method, with a fixed time step of 0.01 seconds;][]{dormand96} to ensure they generated equivalent trajectories under a range of initial conditions (data not shown).

\subsection*{Updating the HV}
HV updating, which could also be referred to as ``path integration proper'', is the process whereby sensory cues relating to velocity are used to update the current HV (see Table 3, HV Updating column). The nature of the updating process will obviously depend on the nature of the HV and the available sensory cues. Nonetheless, all types of HV updating require information about the animal's speed $s$ and also some kind of directional information. The classification of directional cue as allothetic or idiothetic was discussed earlier. In the equational models considered here $\phi$ refers to the animal's compass heading and is most naturally interpreted as an allothetic compass cue, whereas $\dot \phi$ refers to the animal's rotation rate and the natural interpretation is as an idiothetic cue. The HV update equations listed in Table 3 were derived from the geometrically correct GC equations for HV updating (Table 3, GC HV Updating). These were taken from the bicomponent model \citep{mittelstaedt73, mittelstaedt82}, but since they are geometrically exact, they can be considered as mathematical definitions rather than modelling choices. These were then mapped into GP, EC and EP coordinate systems (see Appendix A), such that the four resulting sets of HV update equations were strictly mathematically equivalent, although it should be stressed that they differ with respect to their application (see below).

From the HV update equations, it is simplest to express the updating of geocentric HVs using information relating to the value of $\phi$ and simplest to update egocentric HVs using the value of $\dot \phi$ (Table 3, HV Updating). Interestingly, this is more than a purely mathematical feature (unlike e.g.\ the polynomial approximations used by \citet{muller88}), and appears to be a general property of PI with neurobiological implications: any HV in an egocentric reference frame requires information about rotation (rate), while a geocentric HV requires absolute heading information. This implies allothetic compass cues are more suitable inputs for a geocentric PI system, whereas idiothetic directional cues are more suited to egocentric PI systems. However, this reasoning is not absolute since it is also possible to use an allothetic compass to measure rotations and maintain an egocentric HV (Table 5, Allothetic directional cue column), or to integrate/sum idiothetic rotations to estimate absolute heading for maintaining a geocentric HV (Table 4, Idiothetic directional cue column). Nonetheless, the aforementioned directional-input/HV-coordinate combination seems the most mathematically parsimonious.

Another feature which differentiates the mathematical HV update equations is the variation in the required rate of change in the HV as the animal moves around. As previously noted, the division by $r$ or $r'$ in the polar update equations (Table 3, HV Updating) means that the rate of change of the angular coordinate increases towards infinity the closer the animal gets to home. This is a general feature of polar-like dynamic vectorial HVs. By definition, dynamic vectors have variable angular components which need to be updated during PI. At zero vectorial length, a dynamic vectorial angle is undefined. In contrast, the EC HV update equations show an increase in required rate of change in response to rotation the further the animal is away from home. This is a feature of ES HVs in general. By definition, the egocentric HV is expressed relative to the animal's body axes and therefore must dynamically account for every rotation of the animal itself (unlike geocentric systems). Since the vectors are static, rotation implies remapping of current amplitudes. The longer the HV, the larger the ES vectorial amplitudes, thus requiring a greater absolute change in amplitudes necessary for a given rotation. In fact, only the GC equations have a maximum rate of change, proportional to the animal's maximum forward speed. This is true of GS PI systems generally (as can be see for the two general classes of GS HV update models shown in Appendices C and D). If rapid changes in the HV are associated with increased error accumulation or energetic costs, GS HVs should be favoured by biological PI systems.

A distinguishing characteristic of GC/GS PI systems is that the HV requires only information from speed and directional cues, whereas the other three systems require information about the current state of the HV itself (e.g.\ for EC HVs, to update the $x'$ coordinate requires information about the $y'$ coordinate as well as speed and rotation rate). This suggests that in non-GS PI systems, HV information has to feedback via two pathways to give the baseline HV state, as well as the change of HV state. A GS PI system avoids this complication and should be considered an advantage.

\subsection*{Reading the HV}
Some models of PI describe how to update the HV, but not how to use the HV to control the animal's behaviour. Others have a `readout mechanism', where the HV is assumed to need translating into an alternative reference frame, but do not specify how this will then influence behaviour. This approach is in danger of deferring some portion of the required information processing to an homunculus who looks at the readout and uses it to control behaviour. Here it is assumed instead that the HV must be used to directly control the animal's rotation rate and speed to perform the required PI-related behaviour, the simplest of which is `bee-line' homing. In this way all information processing required of the brain can be made explicit (although some PI-related information processing is beyond the scope of the current paper, such as ephemeris compensation, e.g. \citet{wehner93a}). The homing equations used here were taken from the GC Mittelstaedt model \citep{mittelstaedt00} (see Table 3, GC Homing), and were then mapped into all four standard coordinate systems (Table 3, Homing column) (see Appendix A). This homing model can be seen as more of an arbitrary modelling choice than the HV update equations, since there is no single geometrically correct way to control motion to produce homing. The Mittelstaedt model is arguably the simplest possible, and is easiest to understand in EC form (although originally given in GC): in EC the turning rate equation may be read as `turn left while home is left, turn right while home is right' and the speed equation may be read 'go forwards while home is in front, go backwards while home is behind'. The equations will be considered here as generic steering equations, rather than a detailed biologically realistic model. The equations show that egocentric HVs contain information in a form directly usable for homing (for this simple model at least, and probably for homing-related behaviour in general), whereas geocentric HVs need additional information from the compass and are effectively dynamically translated into egocentric terms before being used for homing.

\subsection*{Leaky Integrator Model of Desert Ant Systematic Navigation Errors}
Several studies have addressed the phenomenon of systematic navigation errors in animals navigating by PI, particularly in desert ants \textit{Cataglyphis fortis} \citep{maurer95, muller88, sommer04}. In the experiments reported by \citet{muller88} the ants travelled outwards from the nest along a two-legged L-shaped channel to the feeder, and were then released to perform homing in a distant test field. In addition to the expected scatter attributed to unbiased random errors, on average the ants homed in a direction different from the geometrically correct homing direction. The deviations from the expected homing direction varied systematically with the angle of the outward L-shaped channels. The M\"uller-Wehner model \textit{ibid.}\ could explain these errors by assuming that the method used by the ant to update its HV during the outward portion of the journey was different from the geometrically correct method. Therefore this model, as well as subsequent attempts to explain these results \citep{hartmann95, merkle06, vickerstaff05, wittmann95}, attributed the homing behaviour to a kind of systematic error that the ants made when updating their HV. Interpretations of the M\"uller-Wehner error should take account of the fact that they were observed in specific experimental settings and analysed in very specific ways, and thus may not relate to the ant's natural PI behaviour in a straightforward manner. In addition, to date, alternative explanations have not been rigorously excluded. Ant navigation is more than PI, and it is plausible that an optimal navigation strategy incorporates PI but does not follow its output exactly. For example the Australian desert ant \textit{Melophorus bagoti} travels only half the homing distance indicated by its PI system under some conditions \citep{narendra07a}. Optimal navigation may also not always entail travelling directly towards the intended target, particularly when additional cues are available \citep{steck09, wolf05}, yet the PI system could still be geometrically exact \textit{per se}. It is also possible that unbiased random errors at some level in an otherwise geometrically exact PI system may produce biased systematic errors in the observed behaviour.

\citet{muller88} provided the first model of systematic errors in desert ants based on approximating the sine and cosine functions, \citet{hartmann95} later produced a neural network model based on the first model, while \citet{wittmann95} provided an alternative mechanism based on a simple time delay, \citet{vickerstaff05} applied a leaky integrator model using a GC HV and \citet{merkle06} undertook a systematic study of several candidate models including leaky integrators, using EC HVs. All of these models sought to explain the PI errors from the L-shaped channels described above, with \citet{vickerstaff05} and \citet{merkle06} independently proposing the same leaky integrator mechanism. In more recent experiments, \citet{sommer04} observed a different kind of systematic error in desert ants, that of underestimating the length of long straight journeys, and examined several candidate models again including a leaky integrator \citep[][introduced a similar model for a human navigation task]{mittelstaedt91}. \citet{wittmann95} proposed that underestimating the return distance, or at least beginning to search for the nest before running off the entire length of the HV, may have the selective advantage of increasing the likelihood of encountering familiar landmarks to guide homing. Recent work in Australian desert ants appears to provide supporting evidence for these ideas \citep{narendra07a, narendra07b}. \citet{merkle06b} also show that, in \textit{Cataglyphis fortis}, depending on the degree of the ant's uncertainty about the home location, systematic search behaviour does not necessarily start after the expected homing distance has been run off.

This section uses leaky integration as a model (see also Appendix B) of systematic navigation errors in order to compare the four coordinate systems under conditions where HV updating is no longer geometrically correct. It will be introduced in GC form first (Table 3, GC Systematic Errors), but like the HV updating and homing models, mathematically equivalent equations for the three other standard coordinate systems are given (Table 3, remainder of Errors column). The GC form assumes that the geometrically correct sensory input values, $s \cos \phi$ and $s \sin \phi$, are available as inputs, but that the integrator itself is `leaky' and decays over time as a first order process with rate constant $k_D$. Hence HV updating is no longer according to the geometrically correct formulae, and the animal could be considered as gradually forgetting earlier parts of its journey. Translation into EC (Table 3 EC Systematic Errors) reveals that, in order to exactly reproduce the behaviour of a GC PI system with leaky integrators, it is sufficient to make the EC HV coordinates behave as leaky integrators with exactly the same decay rate. This result shows that the GC leaky integrator model of \citet{vickerstaff05} and the EC model of \citet{merkle06} are mathematically exactly equivalent (at least in the absence of noise), and predict precisely the same errors for all possible journeys. The translations from GC and EC into GP and EP show that the decay can be considered to directly affect just the magnitude component of the HV since the decay constant only appears in the equations for $r$ and $r'$ and not for $\theta$ or $\theta'$. However, it should be emphasised that the HV of a non-straight journey can still end up pointing in the wrong direction, since a later step is effectively represented on a larger scale than an earlier step of the same physical length, leading to a distortion of the HV.

It is perhaps surprising that deviating from the geometrically exact HV update equations in one coordinate system still yields concise, understandable equations in the other reference frames. In this case the leaky integrator concept maps easily into all four standard coordinate systems. This is probably not the case for all conceivable deviations from geometrical HV updating, but the present result serves as an existence proof that such equivalences do exist and may frustrate attempts to probe the inner workings of biological PI systems using only behavioural data \citep[cf.][]{maurer95}.

The leaky integrator model will now be fitted to two existing data sets from experiments with desert ants to determine the decay rate parameter \citep[this is a reworking of the same findings presented in][using SI units]{vickerstaff05}. The model provides a good quantitative fit to the two sets of experimental data examined, although a different decay rate is required for each set. To model the systematic homing errors observed after L-shaped \citep{muller88} and long straight \citep{sommer04} foraging journeys it is sufficient to consider journeys made up of a small number of straight sections with uniform speed, where the differential equations have simple solutions without the need for numerical integration. The simplest set of assumptions, which are nevertheless reasonably realistic for \textit{C.\ fortis} ants (T. Merkle, personal communication), are that the animal walks with a constant speed ($s$) during foraging and homing, that it spends no significant time at the feeder before beginning to head home, that the decay rate is constant throughout the entire journey and that homing consists of walking straight towards the current estimated home location with no significant variations in heading or speed. The state of a GC HV after walking along a straight section with compass heading $\phi$ at speed $s$ for a time $t$ will be:

\begin{align}
x_t &= \brak{x_0 - \frac{s \cos \phi}{k_D}}e^{-k_Dt}
      + \frac{s \cos \phi}{k_D}\nonumber \\
y_t &= \brak{y_0 - \frac{s \sin \phi}{k_D}}e^{-k_Dt}
      + \frac{s \sin \phi}{k_D} \nonumber
\end{align}

\noindent where the GC HV state at the start of the section is $(x_0,y_0)$ and the decay rate is $k_D$. Upon reaching the end of the foraging part of the journey the ant will adopt a homing direction determined by the current HV state $(x,y)$ such that it orients directly towards the estimated home location. Note this is a simple direct homing path in a straight line, and is not the homing path predicted by the homing equations previously considered; the homing equations essentially show how the ant may adopt orientation $\phi_{\mathrm{homing}}$ without performing an explicit $\arctan$ calculation
$
\phi_\mathrm{homing} = \mathrm{atan2}(-y,-x) \nonumber
$
\noindent (where $\mathrm{atan2}$ is the 4-quadrant arctangent function as defined in Table 1). Switching for convenience to the (strictly equivalent) EP description, the value of the HV with the animal's orientation at $\phi_\mathrm{homing}$ must be $r' = \sqrt{x^2+y^2}, \theta'=0$ since home is now directly ahead. During homing only the value of $r'$ will change according to the leaky integration model (Table 3 EP Systematic Errors), resulting in only one update equation which is simply $\dot r' = -s - k_D r'$ because $\theta'=0$. The value of $r'$ when homing first starts will overestimate the distance the ant walks before the HV reaches zero due to the ongoing decay process. The homing distance will be:

\begin{align}
-\frac{s}{k_D} \ln\brak{\frac{s}{k_D r'_0 + s}} \nonumber
\end{align}

\noindent where $r'_0$ is the HV magnitude at the beginning of homing. This calculates the distance the animal will have walked before the HV magnitude reaches zero, which is taken as the perceived home location (see Appendix A for the derivations).

\begin{figure}
\centering
\includegraphics[width=10cm]{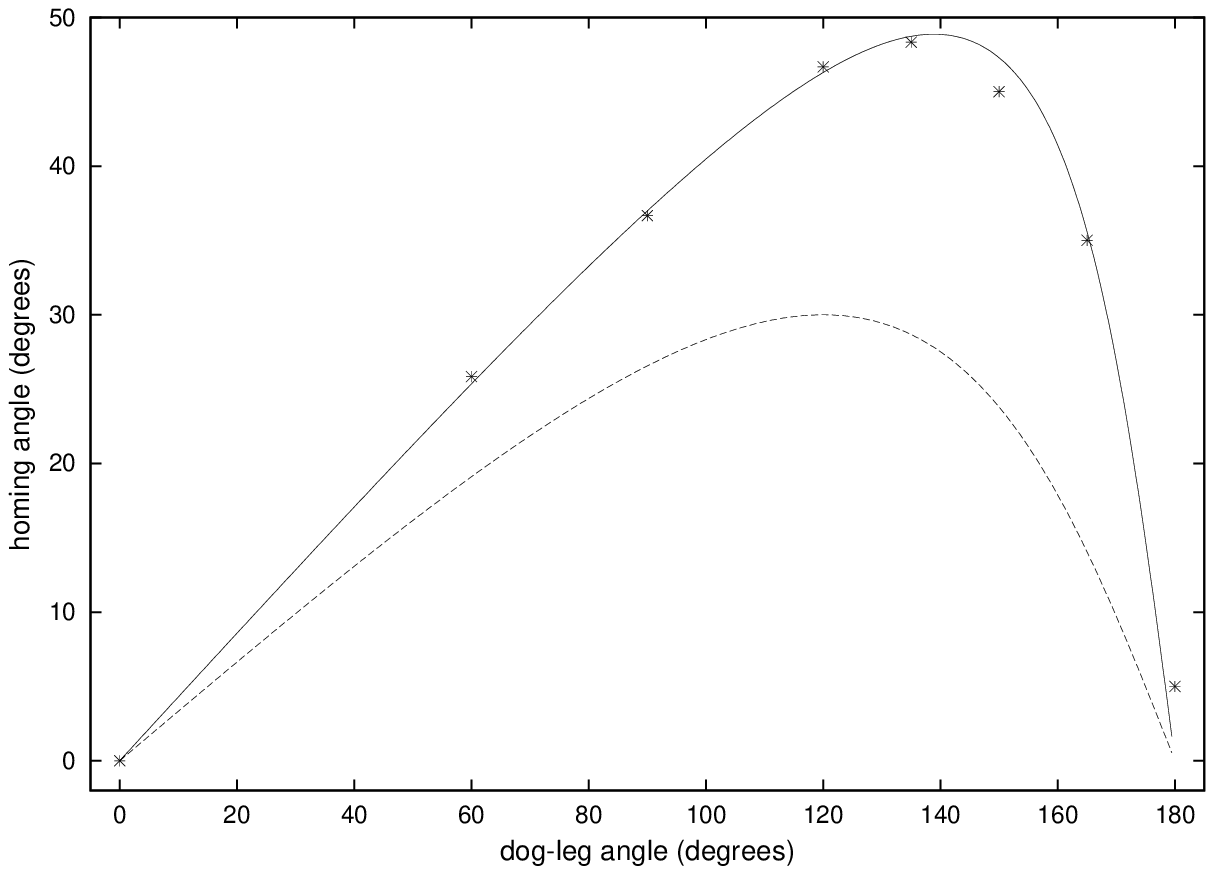}
\caption{The leaky integration model of systematic homing errors fitted to data from the dog-leg experiments of \citet{muller88} where the first section was 10m and the second section 5m in length. x-axis: angle turned by the ant between the first and second section (clockwise from direction of first section). y-axis: homing angle after completing the second section (clockwise from direction opposite that of first section). Solid line: predicted homing angle with a decay rate of $k_D = 0.0185\mathrm{s^{-1}}$ and a walking speed of $0.33\mathrm{ms^{-1}}$. Dashed line: predicted homing angle with decay rate of zero, i.e. the geometrically correct homing angle. Stars: data points from the experiments.}
\label{fig:mullerA}
\end{figure}

\begin{figure}
\centering
\includegraphics[width=10cm]{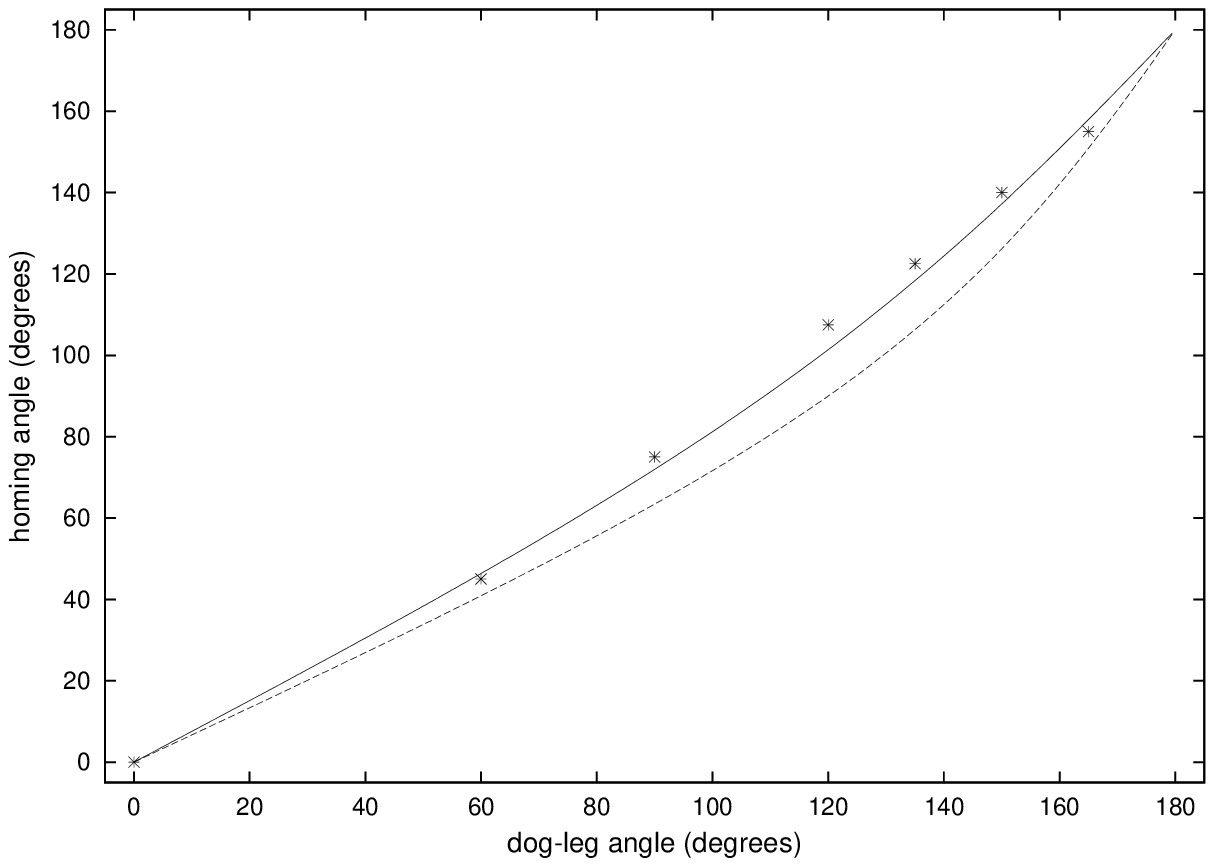}
\caption{The leaky integration model of systematic homing errors fitted to data from the dog-leg experiments of \citet{muller88} where the first section was 5m and the second section 10m in length. x-axis: angle turned by the ant between the first and second section (clockwise from direction of first section). y-axis: homing angle after completing the second section (clockwise from direction opposite that of first section). Solid line: predicted homing angle with a decay rate of $k_D = 0.0185\mathrm{s^{-1}}$ and a walking speed of $0.33\mathrm{ms^{-1}}$. Dashed line: predicted homing angle with decay rate of zero, i.e. the geometrically correct homing angle. Stars: data points from the experiments.}
\label{fig:mullerB}
\end{figure}

\begin{figure}
\centering
\includegraphics[width=10cm]{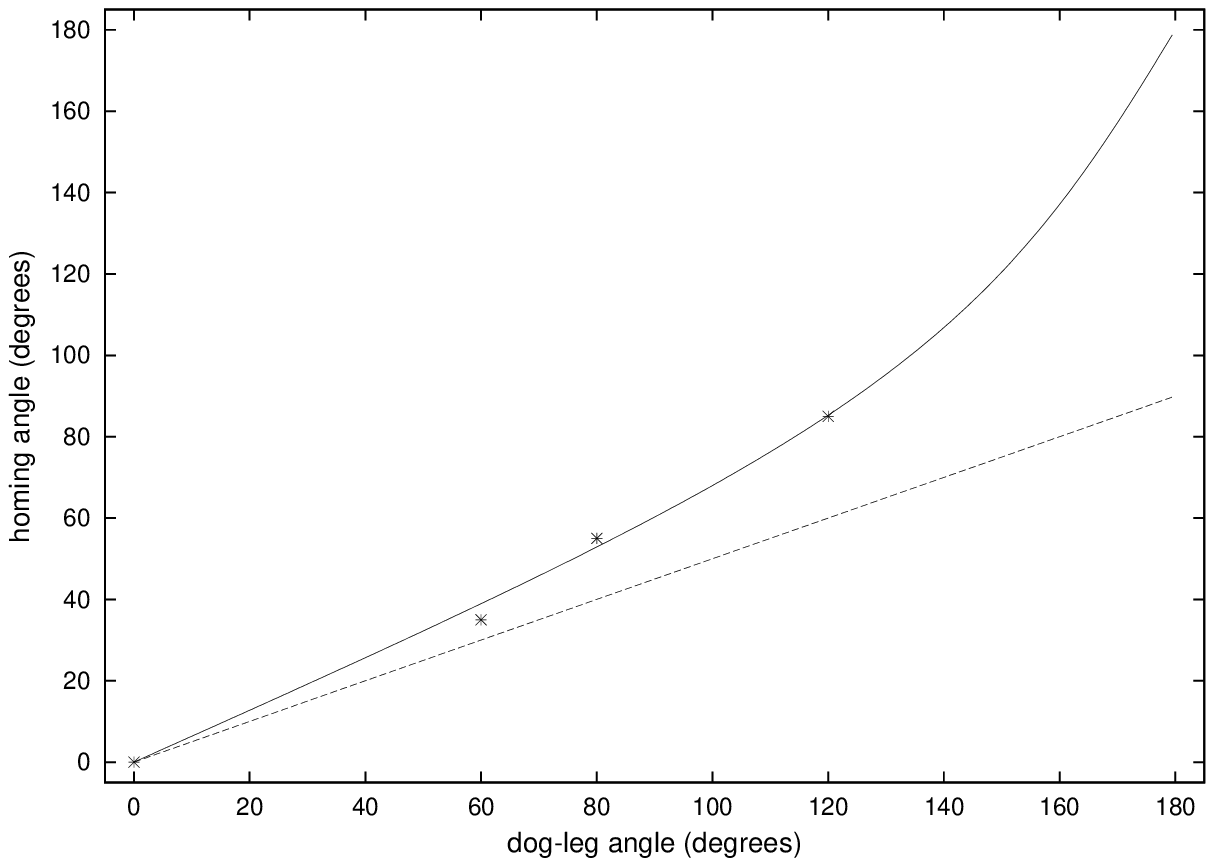}
\caption{The leaky integration model of systematic homing errors fitted to data from the dog-leg experiments of \citet{muller88} where the first section was 10m and the second section 10m in length. x-axis: angle turned by the ant between the first and second section (clockwise from direction of first section). y-axis: homing angle after completing the second section (clockwise from direction opposite that of first section). Solid line: predicted homing angle with a decay rate of $k_D = 0.0185\mathrm{s^{-1}}$ and a walking speed of $0.33\mathrm{ms^{-1}}$. Dashed line: predicted homing angle with decay rate of zero, i.e. the geometrically correct homing angle. Stars: data points from the experiments.}
\label{fig:mullerC}
\end{figure}

\begin{figure}
\centering
\includegraphics[width=10cm]{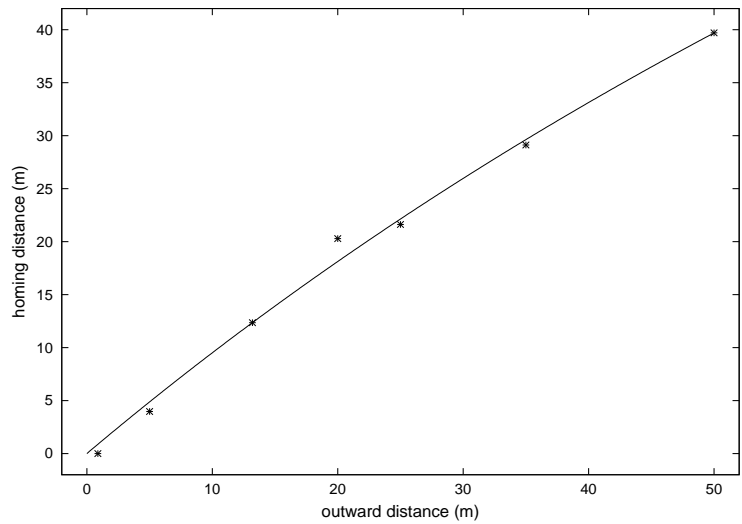}
\caption{The leaky integration model of systematic homing errors fitted to data from the long straight journey experiments of \citet{sommer04}. x-axis: distance walked on outward leg. y-axis: distance walked on return leg. Solid line: predicted homing distance with a decay rate of $k_D = 0.00171\mathrm{s^{-1}}$ and a walking speed of $0.33\mathrm{ms^{-1}}$. The geometrically correct homing distance is equal to the outward distance. Stars: data points from the experiments.}
\label{fig:sommer}
\end{figure}

A zero decay rate yields geometrically correct navigation, regardless of walking speed. \citet{hartmann95} showed figures of the observed homing directions from three different L-shaped outward excursion experiments performed by \citet{muller88}. An average ant walking speed of $0.33\mathrm{ms^{-1}}$ can be estimated from \citet[][Fig.\ 1]{muller88}. Using this value for $s$, a decay rate constant of $k_D = 0.0185\mathrm{s^{-1}}$ is obtained by fitting the leaky integrator model to the homing direction data (data from the three experiments were pooled before fitting $k_D$). Figs \ref{fig:mullerA}-\ref{fig:mullerC} show the predicted homing angles using this $k_D$ value. Assuming the same walking speed, a different decay rate constant ($k_D=0.00171\mathrm{s^{-1}}$) is obtained by fitting the model to the errors observed during long straight journeys reported in \citet{sommer04}. Fig.\ \ref{fig:sommer} shows the predicted homing distances from this decay rate.

\subsection*{Pendulum Model of Desert Ant Systematic Searching Behaviour}
When desert ants reach the end of their HV but do not find the nest entrance, they begin a searching behaviour \citep{wehner81}. This may be a method to compensate for the inevitable errors which accumulate in the HV during foraging excursions, and shows significant flexibility in response to the conditions of the preceding journey. Longer foraging journeys lead to decreased certainly about the nest position, which the ants compensate for by searching a wider area \citep{merkle06b}. The spread of the search also increases the longer an individual search lasts for \citep{muller94, wehner81}. The ants can also learn to bias the shape of their search pattern in response repeated mismatches between the expected and actual nest position \citep{cheng02, wehner02}. Mathematically, the most efficient search pattern would be an ever expanding Archimedean spiral. However, if the ant failed to detect the entrance on its first encounter, it would continue to spiral outwards and would never find it \citep{wehner81}. Hence the real ants do not use this strategy, but perform a series of loops, gradually expanding overtime, but centred on and occasionally returning to the estimated home location. This means the amount of time spent searching per area decreases with distance from the centre, resulting in a search density profile which is approximately symmetrical and bell-shaped. Previous models of searching behaviour in arthropods include \citet{hoffmann83, hoffmann83a, muller94, wehner81}.

To generate a model of searching behaviour suitable for a systematic comparison of the four coordinate systems, it would be convenient to have a simple modification of the homing equations of the Mittelstaedt model which enable it to produce searching. \citet{kim00} have noted how their simple pseudocode-based homing system produced simple repetitive looping behaviour near the home location, and speculated that this might form the basis of a searching behaviour. The neural network model of \citet{vickerstaff05} is very similar to the Mittelstaedt model but also produced searching behaviour which is qualitatively similar to that of desert ants, including a radially symmetric bell-shaped search density. As discussed earlier, the model was produced using a genetic algorithm and was selected to perform PI-mediated homing in the presence of noise, including sensory noise. This led to significant random errors in the HV. As a compensatory strategy the network evolved to perform a searching behaviour upon returning to the estimated home location. The full neural network model is too complex to use here but, fortunately, analysis of the network (R.J.Vickerstaff, unpublished) produced a simple modification of the Mittelstaedt homing equations which generates searching in a manner virtually identical to that of the full neural network model. This section examines the model (Table 3, Searching column), in the original GC reference frame and also following translation into the other three standard PI coordinate systems.

\begin{figure}
\centering
\includegraphics[width=10cm]{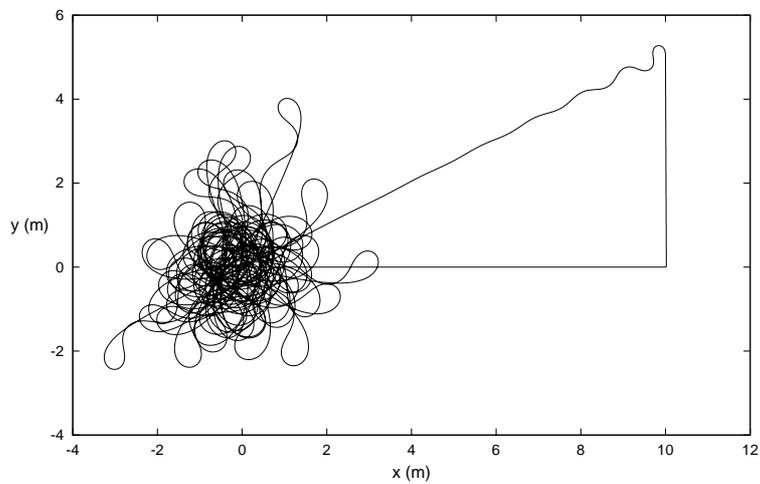}
\caption{Homing trajectory generated by the searching model after an L shaped outwards excursion. The animal begins at location $(0,0)$. The outward journey consists of one leg $10\mathrm{m}$ to the right, followed by $5\mathrm{m}$ upwards. Homing begins immediately after the second leg ends, and is generated by the searching model (including the initial, straight homing leg and the subsequent search pattern centred on the home location). The animal's forward speed is fixed at $1\mathrm{ms^{-1}}$ throughout the entire journey. The model parameters used (see Table 3, Searching) are $s=1$, $k_1=2.7973$ and $k_2=1.308$. The simulation covers 400 seconds. The trajectory was produced by numerical integration of the HV update and searching equations using GC coordinates, using the Runge-Kutta 4th order method with a fixed step size of $0.01$ seconds. No noise is present in the simulation.}
\label{fig:homing}
\end{figure}

\begin{figure}
\centering
\includegraphics[width=10cm]{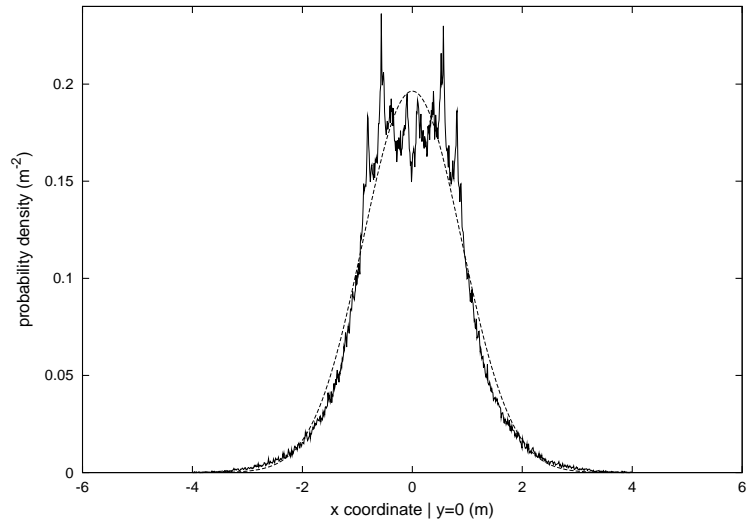}
\caption{Search density profile of a single search pattern lasting for $10^{6}$ seconds, using search parameters $s=1$, $k_1=2.7973$ and $k_2=1.308$ (see Table 3, Searching). The animal was returning from an outward foraging journey as shown in Fig.\ \ref{fig:homing}. The trajectory was produced by numerical integration of the HV update and searching equations using GC coordinates, using the Runge-Kutta 4th order method with a fixed step size of $0.01$ seconds. Search density was recorded (excluding  the outward part of the journey) on an 800 by 800 grid covering an area of 8 by 8 metres centred on the home location. The plot shows a slice through the two dimensional probability density at y=0. For comparison, the dotted line shows a fitted radially symmetrical bivariate normal distribution. The fitted value of $\sigma$ was $0.90028\mathrm{m}$ assuming a correlation of $\rho=0$ (i.e.\ the fitted curve is for a radially symmetrical bivariate normal).}
\label{fig:profile2}
\end{figure}

The model produces straight homing trajectories when the animal is far from home, but close to home a complicated non-repeating, looping search pattern develops (see Fig.\ \ref{fig:homing}). The search patterns still occur even though the noise used during the evolution of the model is removed. Numerical calculation of the maximum Lyapunov exponent of the system \citep{sprott03} shows that the dynamics of the searching pattern are chaotic, indicating that the patterns do not have a simple repetitive structure, even in the absence of external noise. The search density in the presence of noise is a smooth radially symmetrical bell shape (data not shown), in the absence of noise the distribution is still roughly bell shaped (see Fig.\ \ref{fig:profile2}). This is a similar search density profile to that observed in desert ants \citep{wehner81}, but the fine details of the trajectories do not appear to match those of the ants.

While the original neural network operated using a GC HV, the searching dynamics are perhaps easiest to understand in an egocentric reference frame. Since the search profile of the model is radially symmetrical around the home location (as it often is in the real ants) it makes sense to assume that the absolute compass heading at any point during the search is not important \textit{per se}. The egocentric translations of the searching model (Table 3, EC and EP Searching) confirm this since they show analytically that the value of $\phi$ is not needed: the three variables $(x',y',\dot\phi)$ form a self contained (`autonomous') system which encapsulates the complete dynamics of the search model expressed in egocentric terms, referring only to the rotation rate $\dot \phi$ and never to the absolute heading $\phi$:

\begin{align*}
\dot x' &= \dot \phi y' - s\\
\dot y' &= -\dot \phi x'\\
\ddot \phi &= k_1 y' - k_2 \dot \phi\\
\end{align*}

\begin{figure}
\centering
\includegraphics[width=15cm]{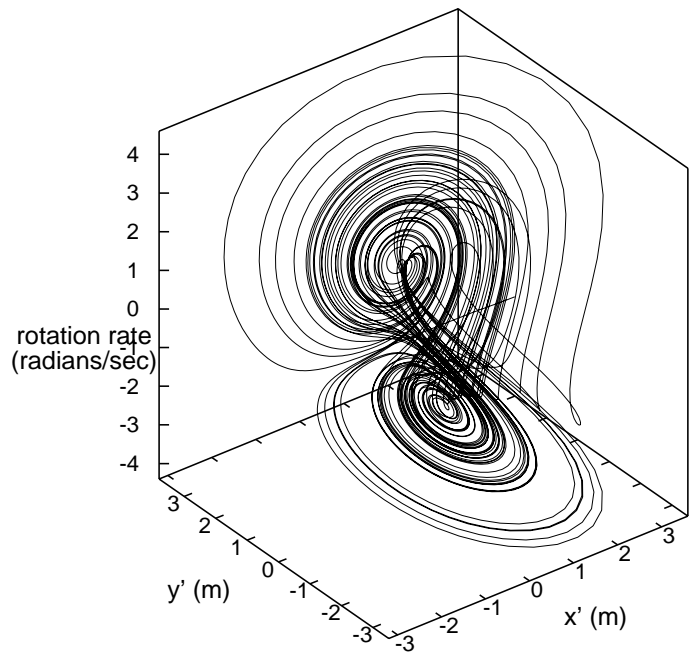}
\caption{Plot of a single search pattern lasting for $400$ seconds, showing the EC HV and rotation rate, using search parameters $s=1$, $k_1=2.7973$ and $k_2=1.308$ (see Table 3, Searching). The animal was returning from an outward foraging journey as shown in Fig.\ \ref{fig:homing}. The trajectory was produced by numerical integration of the HV update and searching equations using EC coordinates, using the Runge-Kutta 4th order method with a fixed step size of $0.01$ seconds. The two main sets of loops in the plot correspond to the parts of the geocentric search when the ant is circling towards the nest in an anticlockwise (positive rotation rate) or a clockwise (negative rotation rate) direction.}
\label{fig:egorotation}
\end{figure}

\noindent Fig.\ \ref{fig:egorotation} shows in EC-rotation space (i.e.\ $(x',y',\dot\phi)$) the search pattern shown in GC space in Fig.\ \ref{fig:homing}. Using parameter values $k_1=2.7973\mathrm{m^{-1}s^{-2}}$, $k_2=1.308\mathrm{s^{-1}}$ and $s=1\mathrm{ms^{-1}}$ the model makes the animal spend half its search time within $1\mathrm{m}$ of the central nest location. Any such egocentric trajectory can be mapped into multiple geocentric searches since it is valid for any absolute orientation. To map an egocentric search trajectory to geocentric space requires choosing an initial $\phi$ value for the start of the time series and applying the appropriate conversion formulae from Table 1 to convert the HV to the corresponding geocentric values. Any value of $\phi$ can be used since the egocentric equations are independent of it. To produce the same search pattern using the geocentric equations requires numerical integration of not three but four variables: $(x,y,\phi,\dot\phi)$. In removing the dependence on absolute orientation the egocentric description employs one less variable but still encapsulates the complete dynamics and makes the underlying ordered structure of the search model more apparent.

As previously noted the EC rotation rate control equation for homing was $\dot \phi = k_\Phi y'$, described as 'when home is to the left turn left, when home is to the right turn right'. The equation for searching controls rotational acceleration rather than rotation rate directly and could be described verbally as 'when home is to the left, rotation tends towards the left, when home is to the right, rotation tends towards the right'. To gain a more detailed understanding of the searching model, consider the EP formulation of the same system (see Table 3, EP HV Updating and Searching):

\begin{align*}
\dot{r'}      &= -s \cos \theta' \\
\dot{\theta'} &= \frac{s}{r'} \sin \theta' - \dot{\phi} \\
\ddot{\phi}   &= k_1 r' \sin \theta' - k_2 \dot{\phi} \\
\end{align*}

\noindent Now consider the special case where forward speed is set to zero ($s=0$), but the animal can still rotate:

\begin{align}
\dot{r'}      &= 0 \nonumber \\
\dot{\theta'} &= - \dot{\phi} \label{eqn:dot_theta_prime} \\
\ddot{\phi}   &= k_1 r' \sin \theta' - k_2 \dot{\phi} \label{eqn:ddot_phi}
\end{align}

\noindent Combining equations \ref{eqn:dot_theta_prime} and \ref{eqn:ddot_phi}:

\begin{align*}
\ddot{\theta'} + k_2 \dot{\theta'}  + k_1 r' \sin \theta' = 0 \\
\end{align*}

\noindent results in an equation exactly homologous to that describing a linearly damped pendulum \citep{strogatz94}, with `gravity' coming from the direction of home (as indicated by the HV) with a force proportional to the distance from home, $r'$, and with $k_2$ as a linear damping constant. When not moving forwards the animal's orientation therefore acts as a damped pendulum which will always come to rest at a stable equilibrium pointing homewards, unless it is perfectly balanced in the direction pointing away from home, but this state is unstable. It is now easier to understand why the model produces direct homing when far from home, but searching behaviour close to home. Strong `gravity' when far away quickly brings the orientation homeward, but when close the attraction is sufficiently weak to allow search loops to form. The searching pattern arises as the complicated interaction of the swings of the pendulum with the changes in the egocentrically perceived homing direction due to forward motion. The values of $k_1$ and $k_2$ must be set correctly to generate search-like behaviour \citep{vickerstaff07}.

The pendulum model's searching equations show that egocentric HVs contain all the necessary information to mediate searching, whereas geocentric HVs must be supplemented with compass information, whereby they are effectively dynamically translated into egocentric terms before being fed into the searching system. This is similar to the simpler homing equations. 

The pendulum model certainly seems to suggest that egocentric PI systems are well suited for mimicking the radially symmetrical Gaussian-like search profiles of desert ants observed under quasi-natural conditions. Whilst symmetrical searching may be simpler, perhaps even more intuitive to quantify in egocentric terms, it does not prove that the underlying PI system uses an egocentric HV. As shown in Table 3, symmetrical search-like behaviour can be generated regardless of the type of HV used. Furthermore, it has been demonstrated that ants may search for their nest asymmetrically if the ant repeatedly experiences an experimentally induced mismatch between its HV and the location of the nest \citep{wehner02}. Asymmetry develops in the searching distribution in the direction which compensates for the mismatch. The latter observation implies that searching behaviour is influenced by geocentric information.

Nonetheless, the preceding analysis suggests that in practice, regardless of the internal PI system, there are advantages in mapping searching and homing trajectories into an egocentric reference frame. This may assist in revealing underlying patterns in the data by removing the absolute orientation from consideration, particularly where the characteristic behaviour is independent of it, and to emphasise the steering control process. The simple pendulum model presented above also demonstrates the utility of chaotic dynamics in generating an efficient, non-repeating search pattern, and that such dynamics may arise spontaneously from simple deterministic rules using nothing more than the HV and current heading. Taken together, these results disprove the argument that complex searching must be a distinct behavioural program to that of early homing. Furthermore, the presence of homing and/or searching behaviour \textit{per se} cannot be used to support or oppose an argument for a particular coordinate system for PI.

\subsection*{Storing Locations with the HV}
Navigation often requires a complex interplay of mechanisms of which PI is one. Although the mathematical analyses presented in this work do not relate directly to other forms of navigation, it is prudent to illustrate the potential relevance of the PI coordinate system to other forms of navigation. Here we briefly consider whether the use of landmarks in combination with PI for navigation may be more readily achieved in certain PI coordinate systems. There is evidence that rodents are able to associate landmark information with the state of the PI system, and use landmark information to correct errors by recalling the associated path integrator state \citep{etienne04}. For this purpose, a geocentric map-like coordinate system possesses distinct computational advantages. Firstly, in a geocentric coordinate system, the coordinates of multiple landmarks may be stored in memory without the need for continual updating as the animal moved around. In contrast, if using egocentric coordinates, landmark coordinates must be updated even if the animal turned on the spot. Secondly, in a map-like PI scheme, landmarks can become linked to the state of the PI system through associative learning in a path-independent manner. In contrast, in some GS HV representations, such as \citet{bernardet08, kim00}, the HV state is partially dependent on the path taken to reach the current location, because the activity levels of neurons are never actively reduced during locomotion. Thirdly, in a map-like scheme, a distinct subset of neural units may be associated to each landmark, reducing the possibility of ambiguity. In contrast, even path-independent GS HV representations, such as \citet{wittmann95}, require every neural unit to be at least partially active at every location in the PI range. Admittedly, polymodal firing fields of grid cells may partly compromise the advantage of the map-like GS scheme in this respect, since they introduce spatial ambiguity in the firing fields, each of which is active at multiple locations in space. Therefore binding a landmark to grid cell activity requires connections to not just one but a population of grid cells \citep{moser08}.

Although it is unclear in arthropods whether the state of the PI system \textit{per se} can associate with landmarks \citep[but see][]{srinivasan97}, there is little doubt that arthropods can store and use vectorial information about goal locations \citep{collett99, frisch67, riley05, wolf00}. If an animal stores a geocentric HV in memory while at a key location, the memory contains sufficient information for returning to the location by PI later. If an egocentric HV is used for this purpose, the animal must also store its compass heading when at the location, otherwise the stored HV does not uniquely specify a geocentric location, but rather a circle of some radius centred on home. Alternatively, the animal must maintain a separate egocentric HV (or rather goal vector) for each goal location and constantly update these while moving. A more general property is seen in Table 1, where transformations between egocentric to/from geocentric reference frames requires the heading $\phi$. In contrast, transforms between static to/from dynamic vectorial representations do not require $\phi$. Therefore, the key piece of information required to translate between egocentric and geocentric space is $\phi$. Such translations will be required in any PI system which uses a combination of geocentric and egocentric vectors due to their different advantages, such as for HV updating versus homing. Without knowing $\phi$, a point in egocentric space corresponds to a circle of possible geocentric locations, and a point in geocentric space corresponds to a circle of possible egocentric locations.

\section*{Conclusions}
It is clear from the literature reviewed in the current work that motivations for using any particular coordinate system for modelling PI varied considerably, particularly in the arthropod literature. Under the novel static/dynamic vectorial classification scheme, it can be seen that a number of models should be considered as variants of the same class, simplifying analyses. For example, a number of PI models have been published in both the arthropod and mammalian literature which used a GC or GS HV \citep{bernardet08, burgess07a, haferlach07, kim00, mcnaughton06, mittelstaedt73, vickerstaff05, wittmann95}. Indeed, one of the earliest PI models ever published, the Mittelstaedt model, was a member of the GS class and was used as the basis for the equational models presented in the paper. Being GC itself, it assumed a pair of sine and cosine input signals from the compass. Although similar signals have been found in nature \citep{lewis98, miller91}, most of the other GS-allothetic models used an array of multiple inputs, including some where the shape of the response function of each input was not sinusoidal. For instance, \citet{wittmann95} assumed a single peaked activation function spread around a circular array of neurons, whereas \citet{bernardet08} assumed a single active neuron indicating direction. Thus, for the special case of a perpendicularly oriented, two valued compass input (a GC representation), strict sinusoidal functions are required for geometrically exact PI, whereas for a larger number the exact shape of the function is less important (see Appendix D). Despite the range of possible neural network realizations of the GS class, the GS HV is fundamentally accumulating distances travelled along a set of geocentrically fixed directions. For the reasons discussed earlier,
(see also Table 7) it possessed relatively few computational disadvantages, but a number of advantages. Amongst these is the fact it is the most suitable coordinate system for HV updating when using an allothetic compass. Given the significant increase in robustness to noise gained from using an allothetic compass for PI \citep{cheung07, cheung08}, this property alone seems to be strong reason to expect some sort of GS HV to be used by biological PI systems where an allothetic compass is available. It is notable that in the mammalian literature, where an abundance of \textit{in vivo} electrophysiological data exists from behaving animals, the few published PI models belong to the GS class.

Which kind of HV is most suitable when only idiothetic directional cues are available? If it is assumed that idiothetic cues are present as information about the animal's rotation rate, then the EC HV update equations \citep{merkle06} appear to be simplest to implement (Table 3, EC HV Updating), followed by EP, although it is also possible to integrate rotation rate to estimate absolute heading for use with either geocentric PI systems (Table 4). However, the disastrous consequences of PI using idiothetic directional cues are now well understood \citep{cheung07, cheung08}, so extreme care must be taken during experimentation and analysis before concluding that any animal is navigating purely by PI in the complete absence of allothetic directional cues.

Separating the HV into a directional and distance component seems to bring unnecessary complications, which apply to both egocentric and geocentric systems. The most obvious drawback from the equations for HV updating is the potential division-by-zero when the animal has a zeroed HV, but more generally the rate of change of the angular part of the HV increases towards infinity the closer the animal is to home. An EC HV has an increasing rate of change the further the animal is from home since the required response of the HV to rotation of the animal is proportional to the distance from home. Given the high rotation rates observed in desert ants, up to 4000$^\circ$ per second in the fastest species \citep{wehner81}, a neural network implementation needs to have a neurophysiogically plausible method of coping with this phenomenon. In contrast, the GC HV has a maximum rate of HV change, limited simply by its maximum forward speed. Furthermore, GC HV coordinates can each be independently updated using information from only speed and directional cues, whereas for GP, EP and EC HVs, correct updating of each HV value requires information about speed, direction and also the other member of the pair of HV values. For these reasons a GC or GS HV representation seems to be a simpler and therefore more robust design, with fewer potential sources of error.

The model of \citet{muller88} employed a GP model to explain systematic homing errors in desert ants, but failed to address the possibility that a similar model using a different coordinate system might produce the same errors. This paper has highlighted how, for the leaky integration model at least, identical results can be obtained regardless of the type of HV used. Whole classes of possible systematic error models may share this property, hence a more rigorous treatment of the subject appears necessary before attempting to use such errors as a probe of the internal workings of biological PI systems.

While egocentric coordinates appear to be necessary and sufficient to control both homing and searching, it is impossible to exclude the possibility that biological PI systems may use a geocentric HV system, dynamically translating the HV into egocentric terms for controlling behaviour. This possibility was highlighted in the earlier description of the pendulum search model where analysis of the behaviour appeared consistent with the idea that the simulated animal was using a rotation rate sensor and an EC or EP HV during search, when in fact it used a GC system throughout. It demonstrated the difficulty of deducing the type of PI system being used based on external observations alone, but still demonstrated the utility of egocentric visualisation of homing and searching data. Thus it is prudent to consider using a variety of reference frames in studying PI behaviour, but care must be taken in drawing conclusions from the analyses.

This paper has largely ignored the effects of noise and random errors on PI. Under this simplifying assumption, we have shown the extent of the mathematical equivalence of alternative coordinate systems for PI, and noted how difficult it may be to determine the internals of a PI system from external behaviour. We introduced a coordinate classification system which adequately covered all exact PI models reviewed. Using this classification scheme, we showed that important properties of PI, especially related to HV updating, behave in similar ways for all members of each extended class. Therefore, this should form a useful foundation for future analyses of PI models, including the study of the effect of neural noise.

Valid arguments were found in favour of and against using each coordinate system. On balance, the current theoretical analysis favours a geocentric and Cartesian or static vectorial HV when an allothetic compass cue is available.  Future work is required to investigate the robustness of the different classes of HV in the face of random noise. This approach may offer the best hope for determining the type of PI coordinate system used in nature. We believe this will be an important step towards an understanding of the neurobiology of animal navigation.


\section*{Appendix A: Derivation of HV Update Equations}
\subsection*{GP from GC}
Starting from the definition of GC HV updating (Table 3), and using the following relations (see Table 1):

\begin{align}
x &= r \cos \theta \nonumber \\
y &= r \sin \theta \label{eqn:georec_to_pol} \tag{A.1} \\
r &= \sqrt{x^2 + y^2} \nonumber
\end{align}

\noindent it can easily be shown that $\dot r = s \cos (\phi - \theta)$ by differentiating $r = \sqrt{x^2+y^2}$. Defining $z = x^2+y^2$, such that $r = \sqrt{z}$ we find that $\dot z = 2 x \dot x + 2 y \dot y$ and $\dot r = \frac{1}{2 \sqrt{z}} \dot z$ using the chain rule. Using the relations given by Eqns \ref{eqn:georec_to_pol}, $\dot z$ expands to:

\begin{align*}
\dot z &= 2 \cdot r \cos \theta \cdot s \cos \phi + 2 \cdot r sin \theta \cdot s \sin \phi\\
\dot z &= 2rs (\cos \theta \cos \phi + \sin \theta \sin \phi)\\
\dot z &= 2rs \cos (\theta - \phi)\\
\dot z &= 2rs \cos (\phi - \theta)\\
\dot r &= \frac{1}{2 \sqrt{z}} 2rs \cos (\phi - \theta)\\
\dot r &= \frac{1}{2 r} 2rs \cos (\phi - \theta)\\
\dot r &= s \cos (\phi - \theta)
\end{align*}

\noindent Next we show that $\dot \theta = \frac{s}{r} \sin (\phi - \theta)$ by differentiating $x = r \cos \theta$, giving $\dot x = \dot r \cos \theta - r \dot \theta \sin \theta$ by the product rule. Rewriting in terms of $r \dot \theta \sin \theta$ gives:

\begin{align*}
r \dot \theta \sin \theta &= \dot r \cos \theta - \dot x\\
r \dot \theta \sin \theta &= s \cos (\phi - \theta) \cos \theta - \dot x\\
r \dot \theta \sin \theta  &= s \cos (\phi - \theta) \cos \theta - s \cos \phi\\
\frac{r}{s} \dot \theta \sin \theta &= \cos (\phi - \theta) \cos \theta -\cos \phi\\
\frac{r}{s} \dot \theta \sin \theta &= (\cos \phi \cos \theta + \sin \phi \sin \theta) \cos \theta -\cos \phi\\
\frac{r}{s} \dot \theta \sin \theta &= \cos \phi \cos^2 \theta + \sin \phi \sin \theta \cos \theta -\cos \phi\\
\frac{r}{s} \dot \theta \sin \theta &= \cos \phi (\cos^2 \theta - 1) + \sin \phi \sin \theta \cos \theta\\
\frac{r}{s} \dot \theta \sin \theta &= \sin \phi \sin \theta \cos \theta -\cos \phi \sin^2 \theta\\
\dot \theta &= \frac{s}{r} (\sin \phi \cos \theta - \cos \phi \sin \theta)\\
\dot \theta &= \frac{s}{r} \sin (\phi - \theta)
\end{align*}

\subsection*{EC from GC}
If $\phi=0$, the direction of the $x$ and $x'$ axes are aligned, and those of the $y$ and $y'$ axes are also aligned, giving the relation $x' = -x$ and $y' = -y$. Rotation of the agent ACW by $\phi$ radians rotates the egocentric axes ACW relative to the geocentric axes, and therefore has the effect of rotating points in egocentric space CW by $\phi$ radians. Therefore $(x',y')$ can be obtained by rotating $(-x,-y)$ CW by $\phi$ radians, or ACW by $-\phi$ radians (from the standard matrix operation of rotation anticlockwise):

\begin{align*}
x' &= (-x) \cos (-\phi) - (-y) \sin (-\phi)\\
y' &= (-x) \sin (-\phi) + (-y) \cos (-\phi)\\
x' &= -x \cos \phi - y \sin \phi\\
y' &= x \sin \phi - y \cos \phi
\end{align*}

\noindent Expressed in reverse $(x,y)$ is obtained by rotating $(-x',-y')$ ACW by $\phi$:

\begin{align*}
x &= (-x') \cos (\phi) - (-y') \sin (\phi)\\
y &= (-x') \sin (\phi) + (-y') \cos (\phi)\\
x &= -x' \cos \phi + y' \sin \phi\\
y &= -x' \sin \phi - y' \cos \phi
\end{align*}

\noindent Differentiating $x'$, using the product rule, and substituting for $\dot x$ and $\dot y$:

\begin{align*}
\dot x' &= x \dot \phi \sin \phi - \dot x \cos \phi - y \dot \phi \cos \phi - \dot y \sin \phi\\
\dot x' &= x \dot \phi \sin \phi - (s \cos \phi) \cos \phi - y \dot \phi \cos \phi - (s \sin \phi) \sin \phi\\
\dot x' &= x \dot \phi \sin \phi - y \dot \phi \cos \phi - s (\cos^2 \phi + \sin^2 \phi)\\
\dot x' &= x \dot \phi \sin \phi - y \dot \phi \cos \phi - s \\
\dot x' &= \dot \phi (x \sin \phi - y \cos \phi) - s \\
\dot x' &= \dot \phi y' - s
\end{align*}

\noindent Differentiating $y'$, using the product rule, and substituting for $\dot x$ and $\dot y$:

\begin{align*}
\dot y' &= x \dot \phi \cos \phi + \dot x \sin \phi + y \dot \phi \sin \phi - \dot y \cos \phi\\
\dot y' &= x \dot \phi \cos \phi + (s \cos \phi) \sin \phi + y \dot \phi \sin \phi - (s \sin \phi) \cos \phi\\
\dot y' &= x \dot \phi \cos \phi + y \dot \phi \sin \phi + s (\cos \phi \sin \phi - \sin \phi \cos \phi)\\
\dot y' &= x \dot \phi \cos \phi + y \dot \phi \sin \phi\\
\dot y' &= \dot \phi (x \cos \phi + y \sin \phi)\\
\dot y' &= -\dot \phi x'
\end{align*}

\subsection*{EP from GP}
It is clear that $r' = r$ at all times, since both measure the distance between the animal and the home location. If $\phi=0$ the $x$ and $x'$ axes are parallel and it is clear that $\theta' = \theta \pm \pi$. Rotating the animal by $\phi$ ACW will reduce $\theta'$ by $\phi$, but leave $\theta$ unchanged. Therefore, the relationship $\theta' = \theta \pm \pi - \phi$ (or $\theta = \theta' \pm \pi + \phi$ ) holds. Taking $\dot r = s \cos (\phi - \theta)$:

\begin{align*}
\dot r' &= s \cos (\phi - \theta)\\
\dot r' &= s \cos (\phi - (\theta' \pm \pi + \phi))\\
\dot r' &= s \cos (- \theta' \pm \pi)\\
\dot r' &= s \cos (\theta' \pm \pi)\\
\dot r' &= -s \cos \theta'
\end{align*}

\noindent Taking $\theta' = \theta \pm \pi - \phi$:

\begin{align*}
\dot \theta' &= \dot \theta - \dot \phi \\
\dot \theta' &= \frac{s}{r}\sin(\phi-\theta) - \dot \phi \\
\dot \theta' &= \frac{s}{r'}\sin(\phi-\theta) - \dot \phi \\
\dot \theta' &= \frac{s}{r'}\sin(\phi-(\theta' \pm \pi + \phi)) - \dot \phi \\
\dot \theta' &= \frac{s}{r'}\sin(-\theta' \pm \pi ) - \dot \phi \\
\dot \theta' &= \frac{s}{r'}\sin \theta' - \dot \phi
\end{align*}

\section*{Appendix B: Leaky integrator model}
\subsection*{Motion with a Fixed Velocity}

\begin{align*}
\dot x = \frac{dx}{dt} = s \cos \phi - k_D x
& \utf \int_{x_0}^{x_t} \frac{dx}{-s \cos \phi/k_D + x} = - k_D \int_0^t dt \\
& \utf \ln\brak{-s \cos \phi/k_D + x} \vert_{x_0}^{x_t} = - k_D t \\
& \utf x_t = s \cos \phi / k_D + \brak{x_0 - s \cos \phi / k_D} e^{-k_D t}
\end{align*}

\begin{align*}
\dot y = \frac{dy}{dt} = s \sin \phi - k_D y
& \utf \int_{y_0}^{y_t} \frac{dy}{-s \sin \phi/k_D + y} = - k_D \int_0^t dt \\
& \utf \ln\brak{-s \sin \phi/k_D + y} \vert_{y_0}^{y_t} = - k_D t \\
& \utf y_t = s \sin \phi / k_D + \brak{y_0 - s \sin \phi / k_D} e^{-k_D t}
\end{align*}

\subsection*{Direct homing}

\begin{align*}
\frac{dr'}{dt} = -s - k_D r'
& \utf \int_{r'_0}^0 \frac{dr'}{s / k_D + r'} = -k_D \int_0^t dt \\
& \utf \ln \brak{s / k_D + r'} \vert_{r'_0}^0 = -k_D t \\
& \utf s / k_D = \brak{s / k_D + r'_0} e^{-k_D t} \\
& \utf t = -\frac{1}{k_D}\ln\brak{\frac{s}{s + k_D r'_0}} \\
r'_{\mathrm{actual}} &= s t = -\frac{s}{k_D}\ln\brak{\frac{s}{s+k_D r'_0}}
\end{align*}

\section*{Appendix C: Exact GS HV Updating Equations}
Any two non-parallel vectors, $\mathbf{u}, \mathbf{v}$, can form the basis of a GS coordinate system. Assuming they are both of unitary length, they can be defined by the direction the vectors point, $\theta_u,\theta_v$, giving vectors of the form:

\begin{align*}
\mathbf{u} = \bmx{\cos\theta_u\\ \sin\theta_u} \\
\mathbf{v} = \bmx{\cos\theta_v\\ \sin\theta_v}
\end{align*}

A HV using this system, $(u,v)$, indicates a GC position of:

\begin{align*}
\bmx{x\\y} = u \bmx{\cos\theta_u\\ \sin\theta_u} + v \bmx{\cos\theta_v\\ \sin\theta_v}
\end{align*}

To find the HV update equations for this system the following must be solved (from the GC HV Update equations given in Table 3):

\begin{align*}
\bmx{\dot x\\ \dot y} = \dot u \bmx{\cos\theta_u\\ \sin\theta_u}
                       + \dot v \bmx{\cos\theta_v\\ \sin\theta_v}
                      = s \bmx{\cos \phi\\ \sin \phi}
\end{align*}

Assuming $\dot u$ to take the form $a_u \cos \phi + b_u \sin \phi$ and $\dot v$ to take the form $a_v \cos \phi + b_v \sin \phi$, it is straightforward to show that:

\begin{align*}
a_u = \frac{\sin\theta_v}{\sin(\theta_v-\theta_u)}\\
b_u = -\frac{\cos\theta_v}{\sin(\theta_v-\theta_u)}\\
a_v = \frac{\sin\theta_u}{\sin(\theta_u-\theta_v)}\\
b_v = -\frac{\cos\theta_u}{\sin(\theta_u-\theta_v)}
\end{align*}

Giving the following result:

\begin{align*}
\dot u &= s \frac{\sin(\theta_v-\phi)}{\sin(\theta_v-\theta_u)}\\
\dot v &= s \frac{\sin(\theta_u-\phi)}{\sin(\theta_u-\theta_v)}
\end{align*}

Thus the update equation for each vector is a scaled and phase shifted sinewave. Rather than the maximum rate of change necessarily being in the direction of the coordinate's associated vector, as might have been expected, instead the rate of change is always zero in the direction of the other coordinate's vector (i.e. $\dot u = 0$ when $\phi = \theta_v$ and vice versa). It is clear that the equations cannot be used if $\sin(\theta_v-\theta_u) = 0$, confirming that only a pair of non-parallel static vectors can be used.

To construct a system using $n$ vectors, each vector may form a pair with the previous vector, and a second pair with the following vector (including the last and first vector pairing), giving $n$ pairs in total. For a set of vectors:

\begin{align*}
\bmx{\cos\theta_i\\ \sin\theta_i}
\end{align*}

forming the basis of an $n$ valued HV $(r_1, \cdots, r_n)$, where the GC coordinate is found using:

\begin{align*}
\bmx{x\\ y} = \sum_{i=1}^n\left( r_i \bmx{\cos\theta_i\\ \sin\theta_i} \right)
\end{align*}

the following update equation may be used (where $n+1 \equiv 1$ and $1-1 \equiv n$):

\begin{align*}
\dot r_i = \frac{s}{n}
\left(
\frac{\sin(\theta_{i-1}-\phi)}{\sin(\theta_{i-1}-\theta_i)}
+
\frac{\sin(\theta_{i+1}-\phi)}{\sin(\theta_{i+1}-\theta_i)}
\right)
\end{align*}

This gives a simple, geometrically exact HV updating system for any $n \ge 2$, provided only that no vectors adjacent in the list are parallel.

\section*{Appendix D: Asymptotically Exact GS HV Update Equations}
For GS vectorial representations, a simple updating mechanism can be implemented using a transfer function $f$ which, like a tuning curve, has a variable response which depends on the current heading $\phi$. The HV representation requires a ring of $n$ neural units, each with the same transfer function $f$, but offset by $\frac{2\pi}{n}$ radians from the adjacent one. Thus $f_j(\phi) = f_{j+1} (\phi + \frac{2\pi}{n})$. Then the HV update equation for axis $j$ is $\dot r_j = s f_j(\phi)$.

Clearly, when $n$ = 4 and $f_1 = \cos \phi$  then  $f_2 = \cos (\phi + \frac{\pi}{2}) = \sin \phi$,  $f_3 = \sin (\phi + \frac{\pi}{2}) = -\cos \phi$, and $f_4 = -\cos (\phi + \frac{\pi}{2} ) = -\sin \phi$, and is equivalent to the standard GC coordinate system, but with negative axes considered as distinct representations in their preferred direction.

When $n$ is small, the simple scheme presented gives an exact HV only for special cases of $f$, like the cosine/sine functions when $n$ = 4. However, as $n \rightarrow \infty$, the structure approaches a continuous representation in angular space hence $\lim_{2\pi / n \to 0} f_0(\phi)= f_\theta(\phi + \theta)$. Below it will be shown that the exact shape of function $f$ becomes unimportant for large $n$.

For any unbiased function $f$ e.g.\ symmetrical about $0$, then $\dot r_\theta = s f_\theta(\phi)$. For an unbiased $f$ we have:

\[
s \int_{-\pi}^{\pi} f_0(\theta+\phi) \cos(\theta+\phi) d\theta = w
\]

and

\[
s \int_{-\pi}^{\pi} f_0(\theta+\phi) \sin(\theta+\phi) d\theta = 0
\]

where $w$ is a constant.

\begin{align*}
\dot x(\phi) &= \int_{-\pi}^{\pi} \dot x_\theta \cos \theta d\theta \\
             &= s \int_{-\pi}^{\pi} f_\theta (\phi) \cos \theta d\theta \\
             &= s \int_{-\pi}^{\pi} f_0 (\phi + \theta) \cos \theta d\theta \\
\end{align*}

From the definitions of an unbiased $f$ we have:

\begin{align*}
w &= s \int_{-\pi}^{\pi} f_0 (\phi + \theta) \cos (\phi+\theta) d\theta \\
  &= s \int_{-\pi}^{\pi} f_0 (\phi + \theta) \cos \phi \cos \theta d\theta
   - s \int_{-\pi}^{\pi} f_0 (\phi + \theta) \sin \phi \sin \theta d\theta \\
  &= s \cos \phi \int_{-\pi}^{\pi} f_0 (\phi + \theta) \cos \theta d\theta
   - s \sin \phi \int_{-\pi}^{\pi} f_0 (\phi + \theta) \sin \theta d\theta \\
0 &= s \int_{-\pi}^{\pi} f_0 (\phi + \theta) \sin (\phi+\theta) d\theta \\
  &= s \int_{-\pi}^{\pi} f_0 (\phi + \theta) \sin \phi \cos \theta d\theta
   + s \int_{-\pi}^{\pi} f_0 (\phi + \theta) \cos \phi \sin \theta d\theta \\
  &= s \sin \phi \int_{-\pi}^{\pi} f_0 (\phi + \theta) \cos \theta d\theta
   + s \cos \phi \int_{-\pi}^{\pi} f_0 (\phi + \theta) \sin \theta d\theta \\
  &= s \tan \phi \sin \phi \int_{-\pi}^{\pi} f_0 (\phi + \theta) \cos \theta d\theta
   + s \sin \phi \int_{-\pi}^{\pi} f_0 (\phi + \theta) \sin \theta d\theta \\
\end{align*}

Summing the two results we get:

\begin{align*}
w + 0 &= s \brak{\cos \phi + \tan \phi \sin \phi}
        \int_{-\pi}^{\pi} f_0 (\phi + \theta) \cos \theta d\theta\\
\therefore w \cos \phi &= s \brak{\cos^2 \phi + \sin^2 \phi}
           \int_{-\pi}^{\pi} f_0 (\phi + \theta) \cos \theta d\theta\\
   &= s \int_{-\pi}^{\pi} f_0 (\phi + \theta) \cos \theta d\theta\\
   &= \dot x(\phi) \\
\end{align*}

Similarly, $\dot y (\phi) = w \sin \phi$. Therefore as $n \rightarrow \infty$, the simple GS HV update equation $\dot r_\theta = s f_\theta(\phi)$ results in the same HV representation as the standard GC equational model. However, in the general GS case, $f$ may be of any form as long as it is unbiased.

These two methods of updating a GS HV prove that varying the number of static vectors cannot be assumed to yield different PI behaviours, since they can be made to reproduce exactly a GC HV. At least two sets of update equations exist (shown above) which make all noise-free static vectorial representations equivalent.

\section*{Role of the funding source}
RJV was supported by the New Zealand Marsden Fund grant UOC0507. AC was supported by an ARC/NHMRC Thinking Systems grant. The funding providers played no role in the design of this study or the decision to publish.

\section*{Contributions}
RJV: Main text, Appendix A, Figures 1 - 8, Tables 1, 3.
AC: Numerous significant contributions to main text, Appendices B - D, Tables 2, 4 - 7.
Both authors have approved the present form of the article.

\section*{Conflict of interest}
We declare that no conflicts of interest exist or have influenced this work.

\section*{Acknowledgements}
The authors wish to thank Thomas Collett, Tobias Merkle and one anonymous reviewer for their help during the preparation of this paper.

\bibliography{main_library}
\bibliographystyle{apalike}

\clearpage

\clearpage

\clearpage

\clearpage

\clearpage

\clearpage

\clearpage

\clearpage

\clearpage
\begin{landscape}
\begin{table}
\small
\begin{tabular}{|>{\footnotesize}l|>{\footnotesize}p{3cm}|>{\footnotesize}p{3cm}|>{\footnotesize}p{3cm}|>{\footnotesize}p{3cm}|}

\hline
From \textbackslash To & GC & GP & EC & EP \\
\hline

GC &
 &
\begin{align}
r = \sqrt{x^2+y^2} \nonumber \\
\theta = \textrm{atan2}(y,x) \nonumber \end{align} &
\begin{align}
x' = -x \cos \phi - y \sin \phi \nonumber \\
y' = x \sin \phi - y \cos \phi \nonumber \end{align} &
\begin{align}
r' = \sqrt{x^2+y^2} \nonumber \\
\theta' = \textrm{atan2}(y,x) \pm \pi - \phi \nonumber \end{align} \\
\hline

GP &
\begin{align}
x = r \cos \theta \nonumber \\
y = r \sin \theta \nonumber \end{align} &
&
\begin{align}
x' = -r \cos (\phi - \theta) \nonumber \\
y' =  r \sin (\phi - \theta) \nonumber \end{align} &
\begin{align}
r' = r \nonumber \\
\theta' = \theta \pm \pi - \phi \nonumber \end{align} \\
\hline

EC &
\begin{align}
x = -x' \cos \phi + y' \sin \phi \nonumber \\
y = -x' \sin \phi - y' \cos \phi \nonumber \end{align} &
\begin{align}
r = \sqrt{x'^2+y'^2} \nonumber \\
\theta = \textrm{atan2}(y',x') \pm \pi + \phi \nonumber \end{align} &
 &
\begin{align}
r' = \sqrt{x'^2+y'^2} \nonumber \\
\theta' = \textrm{atan2}(y',x') \nonumber \end{align} \\
\hline

EP &
\begin{align}
x = -r' \cos (\theta' + \phi) \nonumber \\
y = -r' \sin (\theta' + \phi) \nonumber \end{align} &
\begin{align}
r = r' \nonumber \\
\theta = \theta' \pm \pi + \phi \nonumber \end{align} &
\begin{align}
x' = r' \cos \theta' \nonumber \\
y' = r' \sin \theta' \nonumber \end{align} &
 \\
\hline

\end{tabular}
\normalsize
\caption{Equations for converting home vectors between four standard coordinate systems: geocentric Cartesian (GC), geocentric polar (GP), egocentric Cartesian (EC) and egocentric polar (EP). The equations convert from the system named in the left column to the system named in the top row. $\phi$ indicates the animal's compass heading measured anti-clockwise from the direction of the x axis in radians. $\textrm{atan2}(y,x)$ is defined as  the angle from the x axis to a line from the origin passing through the point $(x,y)$, again measured positive anti-clockwise, and can be calculated using $\arctan\frac{y}{x}+k\pi$, where $k=0$ if $x \ge 0$, $k=1$ if $x \le 0$ and $y \ge 0$, and $k=-1$ if $x \le 0$ and $y \le 0$. }
\label{tab:conversions}
\end{table}
\clearpage
\begin{table}
\begin{tabular}{|>{\scriptsize}l|>{\scriptsize}l|>{\scriptsize}l|}

\hline
Usage & Abbreviation & Meaning \\
\hline
General & PI & Path integration \\
\cline{2-3}
 & HV & Home vector \\
\hline
Standard coordinate systems & GC &  Geocentric Cartesian \\
\cline{2-3}
 & GP &  Geocentric polar \\
\cline{2-3}
 & EC &  Egocentric Cartesian \\
\cline{2-3}
 & EP &  Egocentric polar \\
\hline
Extended coordinate systems & GS &  Geocentric static \\
\cline{2-3}
(vectorial basis representation) & GD &  Geocentric dynamic \\
\cline{2-3}
 & ES &  Egocentric static \\
\cline{2-3}
 & ED &  Egocentric dynamic \\
\hline
Path integration (PI) inputs & $s$ &  Speed $(\mathrm{m s^{-1}})$ \\
\cline{2-3}
 & $\phi$ &  Absolute heading (rad) \\
\hline
Home vector (HV) values & $x,x'$ & Geocentric x distance, egocentric x distance (rostrocaudal axis) (m) \\
\cline{2-3}
 & $y,y'$ & Geocentric y distance, egocentric y distance (left-right axis) (m) \\
\cline{2-3}
 & $r,r'$ & Geocentric radial HV distance, egocentric radial HV distance (m) \\
\cline{2-3}
 & $\theta,\theta'$ & Geocentric HV angle, egocentric HV angle (m) \\
\hline
Constants / coefficients & $k_S$ & Speed coefficient during steering $\mathrm{(s^{-1})}$ \\
\cline{2-3}
 & $k_D$ & Decay rate of first order decay of HV distance representations (leaky
integrator) $\mathrm{(s^{-1})}$ \\
\cline{2-3}
 & $k_\Phi$ & Coefficient of turning during steering $\mathrm{(m^{-1} s^{-1})}$ \\
\cline{2-3}
 & $k_R$ & Search radius within which agent spends half of its time $\mathrm{(m)}$ \\
\cline{2-3}
& $k_1,k_2$ & Weights egocentric y position of home and rotation rate \\
& & to dynamically change homing/search path curvature $\mathrm{(m^{-1}s^{-2}, s^{-1})}$ \\
\hline
\end{tabular}
\caption
{
Abbreviations used in this paper. SI units are included here for convenience but do not affect the form of the equations used in this work.
}
\label{tab:abbrevs}
\end{table}
\clearpage
\begin{table}
\begin{tabular}{|>{\footnotesize}l|>{\footnotesize}p{2cm}|>{\footnotesize}p{2cm}|>{\footnotesize}p{2cm}|>{\footnotesize}p{2cm}|}

\hline
& HV Updating & Systematic Errors & Homing & Searching \\
\hline
GC &
\begin{align}
\dot x = s \cos \phi \nonumber \\
\dot y = s \sin \phi \nonumber \end{align} &
\begin{align}
\dot x = s \cos \phi - k_D x \nonumber \\
\dot y = s \sin \phi - k_D y \nonumber \end{align} &
\begin{align}
\dot \phi = k_\Phi (x \sin \phi - y \cos \phi) \nonumber \\
s = k_S (-x \cos \phi - y \sin \phi) \nonumber \end{align} &
\begin{align}
\ddot \phi = k_1 (x \sin \phi - y \cos \phi) - k_2 \dot \phi \nonumber
\end{align} \\
\hline
GP &
\begin{align}
\dot{r} = s \cos(\phi - \theta) \nonumber\\
\dot{\theta} = \frac{s}{r} \sin(\phi - \theta) \nonumber \end{align} &
\begin{align}
\dot r = s \cos (\phi - \theta) - k_D r \nonumber \\
\dot \theta = \frac{s}{r} \sin (\phi - \theta) \nonumber \end{align} &
\begin{align}
\dot{\phi} = k_\Phi r \sin (\phi - \theta) \nonumber \\
s = -k_S r \cos(\phi - \theta) \nonumber \end{align} &
\begin{align}
\ddot{\phi} = k_1 r \sin (\phi - \theta) - k_2 \dot{\phi} \nonumber
\end{align} \\
\hline
EC &
\begin{align}
\dot x' = \dot \phi y' - s \nonumber \\
\dot y' = -\dot \phi x' \nonumber \end{align} &
\begin{align}
\dot x' = \dot \phi y' - s - k_D x' \nonumber \\
\dot y' = -\dot \phi x' - k_D y' \nonumber \end{align} &
\begin{align}
\dot{\phi} = k_\Phi y' \nonumber \\
s = k_S x' \nonumber \end{align} &
\begin{align}
\ddot{\phi} = k_1 y' - k_2 \dot{\phi} \nonumber
\end{align} \\
\hline
EP &
\begin{align}
\dot{r'} = -s \cos \theta' \nonumber \\
\dot{\theta'} = \frac{s}{r'} \sin \theta' - \dot{\phi} \nonumber \end{align} &
\begin{align}
\dot r' = - s \cos \theta' - k_D r' \nonumber \\
\dot \theta' = \frac{s}{r'} \sin \theta' - \dot \phi \nonumber \end{align} &
\begin{align}
\dot{\phi} = k_\Phi r' \sin \theta' \nonumber \\
s = k_S r' \cos \theta' \nonumber \end{align} &
\begin{align}
\ddot{\phi} = k_1 r' \sin \theta' - k_2 \dot{\phi} \nonumber
\end{align} \\
\hline

\end{tabular}
\caption
{
Continuous-time equations in each of the standard coordinate systems describing HV updating, systematic PI errors, homing and searching. Dot notation indicates the first derivative with respect to time (for example $\dot x$ indicates $\frac{dx}{dt}$), two dots indicates the second derivative with respect to time. Symbols used: $(x,y)$, $(r,\theta)$, $(x',y')$ and $(r',\theta')$ are the GC, GP, EC and EP home vector respectively; $s \mathrm{(m\: s^{-1})}$, the animal's forward speed; $\phi$, the animal's compass heading measured anticlockwise from the x axis direction in radians; $k_D \mathrm{(s^{-1})}$, a positive decay rate; $k_\Phi \mathrm{(s^{-1} m^{-1})}$, a positive turning rate coefficient; $k_S \mathrm{(s^{-1})}$, a positive forward speed coefficient;
$k_1 \mathrm{(s^{-2} m^{-1})}, k_2 \mathrm{(s^{-1})}$, two positive constants controlling searching behaviour. Forward speed is held at a constant value during searching. $s=1, k_1=2.7973, k_2=1.308$ gives a search pattern with a walking speed of $\mathrm{1 \:m\:s^{-1}}$ and a search pattern which spends approximately half of the time within a radius of $\mathrm{1\: m}$ of the origin. To scale to arbitrary walking speed $s$ and search radius $k_R \mathrm{(m)}$ use $k_1=2.7973 \frac{s^2}{k_R^3}, k_2=1.308 \frac{s}{k_R}$.
}
\label{tab:all_models}
\end{table}
\clearpage

\begin{table}
\begin{tabular}{|>{\scriptsize}l|>{\scriptsize}p{4cm}|>{\scriptsize}p{4cm}|}

\hline
& Allothetic directional cue & Idiothetic directional cue \\
\hline

GC & & \\ Continuous &
\begin{align}
\dot x = s \cos \phi \nonumber \\
\dot y = s \sin \phi \nonumber \end{align} &
\begin{align}
\dot x = s \cos \brak{\int \dot \phi dt} \nonumber \\
\dot y = s \sin \brak{\int \dot \phi dt} \nonumber \end{align} \\
\hline

GC & & \\ Discrete &
\begin{align}
x\tdt - x_{t} = s\tdt \cos \phi\tdt \nonumber \\
y\tdt - y_{t} = s\tdt \sin \phi\tdt \nonumber \end{align} &
\begin{align}
x\tdt - x_{t} = s\tdt \cos \brak{\sum \Delta \phi} \nonumber \\
y\tdt - y_{t} = s\tdt \sin \brak{\sum \Delta \phi} \nonumber \end{align} \\
\hline

GP & & \\ Continuous &
\begin{align}
\dot{r} = s \cos\brak{\phi - \theta} \nonumber\\
\dot{\theta} = \frac{s}{r} \sin\brak{\phi - \theta} \nonumber \end{align} &
\begin{align}
\dot{r} = s \cos\brak{\brak{\int \dot \phi dt} - \theta} \nonumber\\
\dot{\theta} = \frac{s}{r} \sin\brak{\brak{\int \dot \phi dt} - \theta} \nonumber \end{align} \\
\hline

GP & & \\ Discrete &
\begin{align}
r\tdt - r_t \approx s\tdt \cos\brak{\phi\tdt - \theta_t} \nonumber\\
\theta\tdt - \theta_t \approx \frac{s\tdt}{r_t} \sin\brak{\phi\tdt - \theta_t} \nonumber\\
r\tdt = \sqrt{r_t^2 + s\tdt^2 + 2 r_t s\tdt \cos\brak{\phi\tdt - \theta_t}} \nonumber\\
\theta\tdt - \theta_t = \sin^{-1}\brak{ \frac{s\tdt}{r\tdt} \sin\brak{\phi\tdt - \phi_t}} \nonumber \end{align} &
\begin{align}
r\tdt - r_t \approx s\tdt \cos\brak{\sum \Delta \phi - \theta_t} \nonumber\\
\theta\tdt - \theta_t \approx \frac{s\tdt}{r_t} \sin\brak{\sum \Delta \phi - \theta_t} \nonumber\\
r\tdt = \sqrt{r_t^2 + s\tdt^2 + 2 r_t s\tdt \cos\brak{\Delta \phi\tdt}} \nonumber\\
\theta\tdt - \theta_t = \sin^{-1}\brak{ \frac{s\tdt}{r\tdt} \sin\brak{\Delta \phi\tdt}} \nonumber \end{align} \\
\hline
\end{tabular}
\caption
{
HV updating equations for GC and GP coordinate systems presented in continuous- and discrete-time, and for the case of allothetic or idiothetic directional cues. Symbol conventions are as for Table \ref{tab:all_models}. $\Delta \phi\tdt = \phi\tdt - \phi_t $. Using an absolute directional cue (true compass), rotation rate can only be estimate e.g.\ $\pd{\phi}{t} = \lim_{\Delta t \to 0} \frac{\phi\tdt - \phi_t}{\Delta t}$.
}
\label{tab:extended1}
\end{table}
\clearpage

\begin{table}
\begin{tabular}{|>{\scriptsize}l|>{\scriptsize}p{4cm}|>{\scriptsize}p{4cm}|}

\hline
& Allothetic directional cue & Idiothetic directional cue \\
\hline

EC & & \\ Continuous &
\begin{align}
\dot x' = \pd{\phi}{t} y' - s \nonumber \\
\dot y' = -\pd{\phi}{t} x' \nonumber
\end{align} &
\begin{align}
\dot x' = \dot\phi y' - s \nonumber \\
\dot y' = -\dot\phi x' \nonumber
 \end{align} \\
\hline

EC & & \\ Discrete &
\begin{align}
x'\tdt - x'_t \approx \brak{\phi\tdt-\phi_t} y'_t - s\tdt \nonumber \\
y'\tdt - y'_t \approx -\brak{\phi\tdt-\phi_t} x'_t \nonumber \\
x'\tdt = x'_t \cos \brak{\phi\tdt-\phi_t} + y'_t \sin \brak{\phi\tdt-\phi_t} - s\tdt \nonumber \\
y'\tdt = -x'_t \sin \brak{\phi\tdt-\phi_t} + y'_t \cos \brak{\phi\tdt-\phi_t} \nonumber
\end{align} &
\begin{align}
x'\tdt - x'_t \approx \Delta\phi\tdt y'_t - s\tdt \nonumber \\
y'\tdt - y'_t \approx -\Delta\phi\tdt x'_t \nonumber \\
x'\tdt = x'_t \cos \brak{\Delta\phi\tdt} + y'_t \sin \brak{\Delta\phi\tdt} - s\tdt \nonumber \\
y'\tdt = -x'_t \sin \brak{\Delta\phi\tdt} + y'_t \cos \brak{\Delta\phi\tdt} \nonumber
 \end{align} \\
\hline

EP & & \\ Continuous &
\begin{align}
\dot r' = -s \cos \theta' \nonumber \\
\dot \theta' = \frac{s}{r'} \sin \theta' - \pd{\phi}{t} \nonumber
\end{align} &
\begin{align}
\dot r' = -s \cos \theta' \nonumber \\
\dot \theta' = \frac{s}{r'} \sin \theta' - \dot\phi \nonumber
 \end{align} \\
\hline

EP & & \\ Discrete &
\begin{align}
r'\tdt - r'_t \approx -s\tdt \cos \theta'_t \nonumber \\
\theta'\tdt - \theta'_t \approx \frac{s\tdt}{r'_t} \sin \theta'_t - \brak{\phi\tdt-\phi_t} \nonumber \\
r'\tdt = \sqrt{r'^2_t + s\tdt^2 + 2 r'_t s\tdt\cos\brak{\phi\tdt-\phi_t}} \nonumber \\
\theta'\tdt-\theta'_t = \sin^{-1}\brak{\frac{s\tdt}{r'\tdt}\sin\brak{\phi\tdt-\phi_t}} + \pi - \brak{\phi\tdt-\phi_t} \nonumber
\end{align} &
\begin{align}
r'\tdt - r'_t \approx -s\tdt \cos \theta'_t \nonumber \\
\theta'\tdt - \theta'_t \approx \frac{s\tdt}{r'_t} \sin \theta'_t - \Delta\phi\tdt \nonumber \\
r'\tdt = \sqrt{r'^2_t + s\tdt^2 + 2 r'_t s\tdt\cos\brak{\Delta\phi\tdt}} \nonumber \\
\theta'\tdt-\theta'_t = \sin^{-1}\brak{\frac{s\tdt}{r'\tdt}\sin\brak{\Delta\phi\tdt}} + \pi - \Delta\phi\tdt \nonumber
 \end{align} \\
\hline

\end{tabular}
\caption
{
HV updating equations for EC and EP coordinate systems presented in continuous and discrete time, and for the case of allothetic or idiothetic directional cues. Symbol conventions are as for Table \ref{tab:all_models}. $\Delta \phi\tdt = \phi\tdt - \phi_t $. Using an absolute directional cue (true compass), rotation rate can only be estimate e.g.\ $\pd{\phi}{t} = \lim_{\Delta t \to 0} \frac{\phi\tdt - \phi_t}{\Delta t}$.
}
\label{tab:extended2}
\end{table}
\clearpage

\begin{table}
\begin{tabular}{|>{\scriptsize}l|>{\scriptsize}p{4cm}|>{\scriptsize}p{4cm}|}

\hline
& Homing & Searching \\
\hline
GC Discrete &
\begin{align*}
\phi\tdt - \phi_t = k_\Phi \brak{x_t \sin \phi_t - y_t \cos \phi_t} \Delta t \\
s\tdt = -k_S \brak{x_t \cos \phi_t + y_t \sin \phi_t} \Delta t
\end{align*}
&
\begin{align*}
\phi\tdt \brak{1+\frac{k_2 \Delta t}{2}} = k_1 \brak{x_t \sin \phi_t - y_t \cos \phi_t} - \phi\tmdt\brak{1-\frac{k_2 \Delta t}{2}} + 2 \phi_t
\end{align*}
\\
\hline
GP Discrete &
\begin{align*}
\phi\tdt - \phi_t = k_\Phi r_t \sin \brak{\phi_t - \theta_t} \Delta t \\
s\tdt = -k_S r_t \cos \brak{\phi_t - \theta_t} \Delta t
\end{align*}
&
\begin{align*}
\phi\tdt \brak{1+\frac{k_2 \Delta t}{2}} = k_1 r_t \sin \brak{\phi_t - \theta_t} - \phi\tmdt\brak{1-\frac{k_2 \Delta t}{2}} + 2 \phi_t
\end{align*}
\\
\hline
EC Discrete &
\begin{align*}
\phi\tdt-\phi_t = k_\Phi y'_t\Delta t \\
s\tdt = k_S x'_t \Delta t
\end{align*}
&
\begin{align*}
\phi\tdt \brak{1+\frac{k_2 \Delta t}{2}} = k_1 y'_t - \phi\tmdt \brak{1-\frac{k_2 \Delta t}{2}} + 2 \phi_t
\end{align*}
\\
\hline
EP Discrete &
\begin{align*}
\phi\tdt-\phi_t = k_\Phi r'_t \sin \brak{\theta'_t} \Delta t \\
s\tdt = k_S r'_t \cos \brak{\theta'_t} \Delta t
\end{align*}
&
\begin{align*}
\phi\tdt \brak{1+\frac{k_2 \Delta t}{2}} = k_1 r'_t \sin \theta'_t  - \phi\tmdt \brak{1-\frac{k_2 \Delta t}{2}} + 2 \phi_t
\end{align*}
\\

\hline

\end{tabular}
\caption
{
The generic homing equations and pendulum search model expressed as discrete time difference equations. Continuous to discrete approximation: $\ddot \phi \approx \frac{\frac{\Delta \phi\tdt}{\Delta t}-\frac{\Delta \phi_t}{\Delta t}}{\Delta t} \approx \frac{\frac{\phi\tdt-\phi_t}{\Delta t}-\frac{\phi_t-\phi\tmdt}{\Delta t}}{\Delta t} = \frac{\phi\tdt + \phi\tmdt - 2 \phi_t}{\Delta t^2}$
}
\label{tab:extended3}
\end{table}

\clearpage

\begin{table}
\small
\begin{tabular}{|>{\footnotesize}l|>{\footnotesize}c|>{\footnotesize}c|>{\footnotesize}c|>{\footnotesize}c|}

\hline
Coordinate System & GC & GP & EC & EP \\
\hline
Rate of HV change approaches infinity near home & & $\bullet$ & & $\bullet$ \\
\hline
Rate of HV change increases with distance from home & & & $\bullet$ &  \\
\hline
HV updating needed in absence of physical translation & & & $\bullet$ & $\bullet$ \\
\hline
Change in HV depends on current HV state & & $\bullet$ & $\bullet$ & $\bullet$ \\
\hline
Further processing required to use allothetic directional cues & & & $\bullet$ & $\bullet$ \\
\hline
Further processing required to use idiothetic directional cues & $\bullet$ & $\bullet$ & & \\
\hline
HV alone insufficient to memorise and return to a geocentric location & & & $\bullet$& $\bullet$\\
\hline
HV can be used for steering towards home & $\circ$ & $\circ$ & $\circ$ & $\circ$ \\
\hline
Linear transform of HV sufficient for homing & & & $\circ$ & \\
\hline
HV can be used for searching for home (e.g.\ pendulum model)&$\circ$&$\circ$&$\circ$ & $\circ$ \\
\hline
Exact update equations exist & $\circ$ & $\circ$ & $\circ$ & $\circ$ \\
\hline
Can accomodate variable speeds of motion & $\circ$ & $\circ$ & $\circ$ & $\circ$ \\
\hline
Can generate appropriate deviations from geometric PI (e.g.\ leaky integrator) & $\circ$ & $\circ$ & $\circ$ & $\circ$ \\
\hline
\end{tabular}
\caption
{
Comparison of the general properties of the main PI related equational models presented in this paper (see Table \ref{tab:all_models}). Filled circles ($\bullet$), likely disadvantage; open circles ($\circ$), likely advantage.
}
\label{tab:properties}
\end{table}
\end{landscape}

\end{document}